\documentclass[11pt,a4paper]{article}
\pdfoutput=1

\usepackage{jheppub}

\usepackage{epsfig,multicol,bbm}
\usepackage{amsmath,amsfonts,amssymb,longtable,mathtools}
\usepackage{array}
\usepackage{graphicx}
\usepackage{subfig}
\usepackage{enumerate}
\usepackage{color}

\def\bea{\begin{eqnarray}}
\def\eea{\end{eqnarray}}

\input epsf.sty

\usepackage[makeroom]{cancel}
  \usepackage{latexsym}
  \usepackage{epsf}
  \usepackage{amssymb}
  \usepackage{graphicx}
  \usepackage{amsmath}
  \usepackage{amsmath,amssymb,amsthm}
  \usepackage{verbatim}
  \usepackage{hyperref}
\renewcommand{\d}{\textrm{d}}

\def\ni{\noindent}
\def\bi{\begin{itemize}}
\def\ei{\end{itemize}}
\def\be{\begin{equation}}   
\def\ee{\end{equation}}
\def\ben{\begin{equation*}}
\def\een{\end{equation*}}
\def\c1{C_1}
\def\d1{D_1}

\begin{document}
\preprint{MI-TH-1617}
\title{Dark Matter Relics and the Expansion Rate in  Scalar-Tensor  Theories}

\author{Bhaskar Dutta\footnote{dutta@physics.tamu.edu}$^{a}$,}
\affiliation{${}^a$Mitchell Institute for Fundamental Physics and Astronomy, Department of Physics and Astronomy, Texas A\&M University, College Station, TX 77843, USA}
\author{Esteban Jimenez\footnote{este1985@physics.tamu.edu}$^{a}$,}
\author{Ivonne Zavala\footnote{e.i.zavalacarrasco@swansea.ac.uk}$^{b}$}
\affiliation{${}^b$Department of Physics, Swansea University, Singleton Park, Swansea, SA2 8PP, UK}

\abstract{We study the impact of a modified expansion rate on  the dark matter relic  abundance  in a class of  scalar-tensor theories.  The scalar-tensor theories we consider are motivated from string theory constructions, which have  
 conformal as well as disformally coupled matter to the scalar. We investigate the effects of such a conformal coupling to the dark matter relic abundance for a wide range of  initial conditions, masses and cross-sections. We find 
that exploiting all possible   initial conditions, the annihilation cross-section required to satisfy the dark matter content can  differ from  the thermal average cross-section in the standard case. 
We also study the expansion rate in the disformal case and find that physically relevant solutions  require a nontrivial relation between the conformal and disformal functions. 
We study the effects of the disformal coupling in an explicit example where the disformal function is quadratic.}


\maketitle

\section{Introduction}\label{introduction}
 
The most recent cosmological observations support the so called $\Lambda CDM$ model of cosmology.  These observations  provide overwhelming evidence for the
existence of non-baryonic (cold) dark matter  (DM), constituting about $27\%$ of the Universe's energy density budget, while another $\sim 68\%$ is believed to be in the
form of dark energy, which can simply be a cosmological constant, $\Lambda$. The other $\sim 5\%$ being formed by baryonic matter, described by the standard model (SM) of particles.

A  popular framework to understand the origin of DM is the thermal relic scenario. In this scenario, at very early times when the universe was at a very high
temperature, thermal equilibrium was obtained and the number density of DM particles $\chi$ was roughly equal to the number density of photons.
During equilibrium the dark matter number density decayed exponentially as $n_\chi^{eq} \sim e^{-m_\chi/T}$ for a non-relativistic DM candidate, where $m_\chi$ is the mass of the DM
particle $\chi$. As the universe cooled down as it expanded, DM interactions became less frequent and eventually,  the DM interaction rate dropped below
the expansion rate ($\Gamma_\chi < H$). At this point the density number froze-out and the universe was left with a ``relic" of DM particles.  
Therefore, the dependence of the number density at the time of freeze-out is crucial to determine the DM relic abundance. The longer the DM particles remain in 
equilibrium, the lower its density will be at freeze-out and vice-versa. In the standard $\Lambda CDM$ scenario, particle freeze-out happens during the 
radiation era and DM species with weak scale interaction cross-section freeze-out with an abundance that matches the present observed value. The weakness of
the interactions is reflected in the predicted thermally-averaged annihilation cross section, $\langle\sigma v \rangle$, which is around $3.0\times10^{-26}cm^3s^{-1}$. Despite such a small value, the Fermi-LAT and Planck experiments have been exploring upper bounds on $\langle\sigma v \rangle$ (see  \cite{Fermi-LAT:2016uux,Ade:2015xua}). From  observations, it appears that the annihilation cross-section can be smaller than the thermal average value for lower dark matter masses ($\leq 100$ GeV), whereas an annihilation cross-section  larger than the thermal average value can still be allowed for larger DM mass.

If, from future  measurements,  $\langle\sigma v \rangle\neq3.0\times10^{-26}cm^3s^{-1}$ is established, what can we say about the origin of the dark matter? Can it still be the thermal dark matter or do we need non-thermal origin of dark matter? In the case of non-thermal origin, the DM can arise from the decay of a heavy particle, e.g., moduli, and can satisfy the DM content with any value of $\langle\sigma v \rangle$.   The primary motivation of this paper is to find out whether the DM content can still have a thermal origin with larger or smaller $\langle\sigma v \rangle$ by utilising non-standard cosmology.

The phenomenological $\Lambda CDM$ model complemented with the inflationary paradigm to provide the seeds
of large scale structure, is very successful in describing our current universe.  However, the physics describing the universe's evolution from the end of inflation (reheating) to  just before big-bang  nucleosynthesis (BBN) ($t\lesssim 200$ s) remains relatively unconstrained. During this period the universe may have gone through  a ``non-standard" period of expansion, and still be compatible with BBN. If such  modification happened during DM decoupling, the DM freeze-out may
be modified with  measurable consequences for the relic DM abundances. 

Departures from the standard cosmology between reheating and BBN will mainly be a consequence of   a modified expansion rate ($\tilde H$), due to a modification of General Relativity (GR). Such modifications are well motivated by attempts to embed the $\Lambda CDM$ and inflation models into a fundamental theory of gravity and 
particle physics, such as theories with extra-dimensions, supergravity and string theory. Indeed, our main motivation in this paper is to develop further tools that may allow us to connect such fundamental theories with observations. 
In this paper our approach will be mostly phenomenological, but we have in mind a scenario that can be derived in the context of a fundamental theory of gravity, such as string theory. Furthermore, we will be concerned with modifications to the standard picture due to the presence and interactions with scalar fields only.

Modifications to the relic abundances were first discussed by Catena et al.~\cite{Catena} in the context of conformally coupled scalar-tensor theories (ST), such as generalisations of the  Brans-Dicke theory. Further studies on conformally coupled ST models have been performed  in the last years in
\cite{Catena2,Gelmini,RG,WI}  (see also \cite{LMN,Pallis,Salati,AM,IC,MW2,MW3}) with the most recent work of \cite{MW} where it is shown that the boundary conditions used in~\cite{Catena} cannot  have  $\langle\sigma v \rangle\neq3.0\times10^{-26}cm^3s^{-1}$. 
Indeed ST theories where dark matter and dark energy are correlated constitute an attractive way to address the dark matter and dark energy problems via an attraction mechanism towards standard general relativity (GR) \cite{Catena}.

In the context of ST theories, the most general physically consistent relation between two metrics in the presence of a  scalar field, is given by\footnote{In the more general case, $C$ and $D$ can be functions of $X=\frac12 (\partial \phi)^2$ as well. However, we will not consider this case in the present paper. } \cite{Bekenstein}:
\be\label{Dmetric}
\tilde g_{\mu\nu} = C(\phi) g_{\mu\nu} + D(\phi) \partial_\mu \phi \partial_\nu \phi\,.
\ee
The first term in \eqref{Dmetric}  is the well-known conformal transformation which
characterises the Brans-Dicke class of scalar-tensor theories explored in \cite{Catena,Catena2,Gelmini,RG,WI,MW}. However, in reference~\cite{MW}, it is shown that there is  no change in the thermal cross-section arising from the conformally modified metric compared to the standard cosmology after satisfying all the constraints based on the boundary conditions chosen in~\cite{Catena,Catena2}. The second term is the disformal contribution, which is generic in extensions of general relativity. In particular, it arises naturally in D-brane models, as  discussed in \cite{KWZ} in a model of coupled dark matter and dark energy.  In this paper we   revisit the expansion rate modification and impact on the DM relic abundances for the conformal case, providing new interesting results with new boundary conditions to show that $\langle\sigma v \rangle\neq3.0\times10^{-26}cm^3s^{-1}$, while satisfying the DM content. 
We further  discuss  the general modifications to the expansion rate and Boltzmann equation,  due to the disformal coupling and present an explicit  non-trivial example for the case in which the conformal term  in  \eqref{Dmetric} is a monomial.  

The paper is organised as follows. We start in section \ref{Sec:1}  introducing the scalar-tensor theory conformally and disformally coupled to matter. Then, we examine the formulation of this
theory in the Einstein and Jordan frames, comment on their physical interpretation and derive the equations that describe the cosmological evolution of the Universe. Subsequently, after discussing the expansion rate modifications caused by the presence of the conformal and disformally coupled scalar field in section \ref{Sec:2}, we investigate in detail its impact on the dark matter relic abundance by exploring a concrete pure conformal example as well as an example where we also turn on the  disformal contribution. Finally, in section \ref{Sec:3} we conclude.

\section{The scalar-tensor theory set-up}\label{Sec:1}

We are interested in  scalar-tensor theories coupled to  matter   both conformally and disformally \cite{Bekenstein}. 
Our motivation comes  from theories with extra dimensions and in particular string theory compactifications,
where several additional scalar fields appear, from  closed and open string theory sectors of the theory \cite{KWZ}. 
Our approach in this paper nonetheless,  will  be phenomenological and therefore our equations will be  simplified.  
However, we present the more general set-up, which can accommodate   a  realisation from concrete string theory  compactifications in  appendix \ref{App1} and \ref{App2}. 
The action we  want to consider is given by: 
\bea\label{S1}
 S_{EH}=\frac{1}{2\kappa^2} \!\!\int{\!d^4x\sqrt{-g}\,R}
- \!\!\int{\!d^4x\sqrt{- g} \left[\frac{1}{2} (\partial\phi)^2+V(\phi)\right]} 
- \!\!\int{\!d^4x\sqrt{-\tilde g} \,{\cal L}_{M}(\tilde g_{\mu\nu}) } \,. 
\eea
Here the disformally coupled metric is given by
\be\label{gtilde}
\tilde g_{\mu\nu} = C(\phi) g_{\mu\nu} + D(\phi) \partial_\mu \phi \partial_\nu \phi\,,
\ee
and the  inverse by:
\be\label{gtildein}
\tilde g^{\mu\nu} = \frac{1}{C}\left[ g^{\mu\nu} - \frac{D\,\partial^\mu \phi \partial^\nu \phi}{C+D(\partial\phi)^2}\right]\,.
\ee
Moreover, $\kappa^2=M_{P}^{-2}=8\pi G$, but keep in mind that  that $G$ is not in general equal to Newton's constant as measured by e.g.~local experiments. Further, $C(\phi), D(\phi)$ are functions of $\phi$, which can be identified as a conformal and disformal couplings of the scalar to the  metric, respectively (note that the conformal coupling is dimensionless, whereas the disformal one has units of $mass^{-4}$).

 The action in \eqref{S1} is written in the Einstein frame\footnote{In  string theory,  the Einstein frame refers  to the frame in which the dilaton and  graviton degrees of freedom are decoupled, while the string (or Jordan) frame is that in which they are not. Further, the dilaton field as well as all other moduli (scalar) fields not relevant for the cosmological  discussion are stabilised, massive, and are therefore decoupled from the low energy effective theory. 
In the literature of  scalar-tensor theories however,  the Einstein and Jordan frames are identified with respect to the (usually single) scalar field to which gravity is coupled, but such scalar has no particular physical  nor geometrical interpretation.}, which is identified in the literature of  scalar-tensor theories  (including conformal and disformal couplings) with the frame respect to which the scalar field, gravity is coupled. We follow this use and refer to  ``Jordan" or ``disformal frame"  to identify the frame in which dark matter is coupled only to the metric $\tilde g_{\mu\nu}$, rather than to the metric $g_{\mu\nu} $ and a scalar field $\phi$.


The equations of motion obtained from  (\ref{S1}) are:
\be\label{EM}
R_{\mu\nu} -\frac{1}{2}g_{\mu\nu} R = \kappa^2\left(T^\phi_{\mu\nu} + T_{\mu\nu}\right)\,,
\ee
where, in the  frame  relative to $g_{\mu\nu}$, the energy-momentum tensors are defined as
\bea\label{EM1}
T^\phi_{\mu\nu} = -\frac{2}{\sqrt{-g}} \frac{\delta S_{\phi}}{\delta g^{\mu\nu}}  \,,\qquad \quad 
T_{\mu\nu}= -\frac{2}{\sqrt{-g}} \frac{\delta \left(-\sqrt{-\tilde g} \,{\cal L}_{M}\right) }{\delta g^{\mu\nu}}\,,
\eea
and we model the energy-momentum tensor for matter and both dark components as perfect fluids, that is:
\be\label{EM2}
T_{\mu\nu}^i = P_i g_{\mu\nu} +(\rho_i + P_i) u_\mu u_\nu
\ee
where $\rho_i$, $P_i$ are the energy density and pressure for each  fluid $i$ with equation of state $P_i/\rho_i = \omega_i$. For the scalar field, the energy-momentum tensor takes the form:
\be\label{EMphi}
T_{\mu\nu}^{\phi} = - g_{\mu\nu} \left[ \frac{1}{2} (\partial\phi)^2+ V \right] 
+  \partial_\mu\phi \, \partial_\nu \phi\,,
\ee
and one can define the energy density and pressure of the scalar field as:
\bea\label{rhoP}
\rho_\phi =- \frac{1}{2}(\partial\phi)^2 +  V  \,, \qquad P_\phi = - \frac{1}{2}(\partial\phi)^2 - V  \,.
\eea
Finally the equation of motion for the scalar field dark energy becomes:
\bea\label{Eqphi}
&&\hskip-1cm - \nabla_\mu \nabla^\mu \phi \! + V' 
- \frac{T^{\mu\nu}}{2}\!\left[\frac{C'}{C} g_{\mu\nu}  +\frac{D'}{C}\partial_\mu\phi\partial_\nu\phi\right]
+\nabla_\mu \left[\frac{D}{C}T^{\mu\nu} \partial_\nu\phi \right] =0 \,.
\eea
 Due to the nontrivial coupling, the individual conservation equations for the two fluids are modified. However, the conservation equation for the full system is preserved, and given in the usual way by
\be
\nabla_\mu \left(T^{\mu\nu}_{\phi} + T^{\mu\nu}\right) =0\,. 
\ee
Thus using (\ref{EMphi})  and the equation of motion for the scalar field we can write 
\be\label{conserva}
\nabla_\mu T^{\mu\nu}_\phi = Q \,\partial^\nu \phi = - \nabla_\mu T^{\mu\nu}\,,
\ee
where
\be
Q\equiv \nabla_\mu \left[\frac{D}{C} \,T^{\mu\lambda} \,\partial_\lambda \phi\right] - \frac{T^{\mu\nu} }{2} \left[\frac{C'}{C} g_{\mu\nu} +
\frac{D'}{C} \,\partial_\mu\phi \,\partial_\nu\phi\right]\,.
\ee

\bigskip 

In the Jordan, or disformal frame, as defined above,  matter is conserved,
\be\label{conservaD}
\tilde \nabla_\mu \tilde T^{\mu\nu} =0  \,,
\ee
where $\tilde \nabla_\mu$ is the covariant derivative computed with respect to the disformal metric \eqref{gtilde} with  the Christoffel symbols   given by
\bea\label{Chris}
 \tilde \Gamma^\mu_{\alpha\beta} =
\Gamma^\mu_{\alpha\beta} + \frac{C'}{C}\, \delta^\mu_{(\alpha}\partial_{\beta)}\phi 
- \gamma^2 \, \frac{C'}{2C} \,\partial^\mu\phi \, g_{\alpha\beta}  
+\frac{D}{C}\,\gamma^2\,\partial^\mu\phi \left[\!\nabla_\alpha\nabla_\beta \phi + \!
 \left(\frac{D'}{2D} -\!\frac{C'}{C}\right) \!\partial_\alpha\phi \partial_\beta \phi \right]\,, \nonumber\\
\eea
and we have introduced the ``Lorentz factor" $\gamma$ defined as 
\be\label{Lorentz}
\gamma = \frac{1}{\sqrt{1+\frac{D}{C} (\partial\phi)^2}}\,.
\ee
In this frame,  the energy-momentum tensor is defined as 
\be\label{tmunuD}
\tilde T^{\mu\nu}  = \frac{2}{\sqrt{-\tilde g}} \frac{\delta S_{M}}{\delta \tilde g_{\mu\nu} } \,
\ee
and  the disformal energy-momentum tensor can be written as:
\be\label{pfluiD}
\tilde T^{\mu\nu}  = (\tilde \rho +\tilde P) \tilde u^{\mu}  \tilde u^{\mu} + \tilde P\, \tilde g^{\mu\nu}\,,
\ee
where   $\tilde u^{\mu} = C^{-1/2} \gamma \,u^{\mu} $. 
Using \eqref{tmunuD}, we  obtain a relation between the energy momentum tensor in both frames as:
\be\label{tmunuDD}
\tilde T^{\mu\nu} = C^{-3} \gamma \, T^{\mu\nu}\,.
\ee
Further using  \eqref{pfluiD} we  arrive at a relation among the energy densities and pressures in both frames, given by
\be\label{eqrhos}
\tilde \rho = C^{-2} \gamma^{-1} \rho \,, \qquad \tilde P = C^{-2}\gamma\, P,
\ee
and therefore the equations of state in both frames are related by $\tilde \omega = \omega\, \gamma^2$. Note that in the pure conformal case, $D=0$, $\gamma=1$ and therefore $\tilde \omega = \omega $.

\subsection{Cosmological equations}

Consider an homogeneous and isotropic  FRW metric $g_{\mu\nu}$, 
\be
ds^2 = - dt^2 + a(t)^2 dx_i dx^i  \,,
\ee
where $a(t)$ is the scale factor.  In this background, the Einstein and  Klein-Gordon equations  become, respectively
\bea
&& H^2 =\frac{\kappa^2}{3} \left[\rho_\phi +\rho\right]\,, \label{friedmann1}\\
&& \dot H + H^2 = -\frac{\kappa^2}{6}\left[ \rho_\phi+ 3P_\phi +\rho +3 P \right]\,,\label{friedmann2}\\
&& \ddot \phi +3H\dot\phi +  V_{,\phi}+Q_0 =0 \label{kgSimple}\,.
\eea
where, $H= \frac{\dot a}{a}$, dots are derivatives with respect to $t$ and we have denoted $V_{,\phi} \equiv \frac{dV}{d\phi}$. Also the Lorentz factor becomes $$\gamma= (1-D \,\dot\phi^2/C)^{-1/2}.$$
The continuity equations for the scalar field and matter are  given by
\bea 
&&\dot\rho_\phi + 3H(\rho_{\phi}+P_{\phi}) = -Q_0\dot\phi\,, \label{cont}\\
&&\dot\rho + 3H(\rho+P) = Q_0\,\dot\phi\,.\label{cont1}
\eea
where $Q_0$ is given by
\bea
Q_0 = \rho \left[ \frac{D}{C} \,\ddot \phi + \frac{D}{C} \,\dot \phi \left(\!3H + \frac{\dot \rho}{\rho} \right) \!+ \!\left(\!\frac{D_{,\phi}}{2C}-\frac{D}{C}\frac{C_{,\phi}}{C}\!\right) \dot\phi^2 +\frac{C_{,\phi}}{2\,C} (1-3\,\omega)
\right]. \nonumber \\
\eea
Using \eqref{cont1} we can rewrite this in a more compact  and useful form as 
\be\label{Q0}
Q_0 = \rho\left( \frac{\dot \gamma}{\dot \phi\, \gamma} + \frac{C_{,\phi}}{2C}  (1-3\,\omega \,\gamma^2) -3H\omega\,\frac{(\gamma^{2}-1) }{\dot \phi}\right)\,.
\ee

\ni Plugging this into  the  
(non-)conservation equation for dark matter \eqref{cont1},  gives:
\be\label{conservaDM}
\dot \rho + 3H (\rho + P\,\gamma^{2}) = \rho \left[\frac{\dot \gamma}{\gamma} + \frac{C_{,\phi} }{2C} \,\dot\phi\, (1-3\,\omega \gamma^2)\right]\,.
\ee
Using the relations for the physical proper time and the scale factors in the two frames, given by
\be\label{tildea}
\tilde a = C^{1/2} a   \,, \qquad \quad d\tilde \tau = C^{1/2} \gamma^{-1} d\tau\,,
\ee
we can define the disformal-frame Hubble parameter $\tilde H \equiv \frac{d \ln{\tilde a}}{d\tilde \tau}$, as
\be\label{tildeH}
\tilde H = \frac{\gamma}{C^{1/2}}\left[ H + \frac{C_{,\phi}}{2C}\dot \phi \right]\,,
\ee
so that \eqref{conservaD} takes  the standard form in terms of $\tilde H$:
\be\label{tilderho2}
\frac{d{\tilde \rho}}{d\tilde\tau}  +  3 \tilde H( \tilde \rho +\tilde P)  =0 \,.
\ee
Equations \eqref{tildeH} and \eqref{tilderho2} give the background evolution equations for the modified expansion rate and  matter's density evolution. 

\subsection{Master equations} \label{ME}

In order to solve the cosmological equations, it is convenient to replace time derivatives with derivatives with respect to the number of e-folds $N$, defined as $ N = \ln a/a_0$  and define  $\lambda  = \frac{V}{ \rho}(=\frac{\tilde V}{\tilde \rho})$.
With these definitions, we can rewrite the Friedmann equation \eqref{friedmann1} and $Q_0$ as:
\bea 
H^2 &=& \frac{\kappa^2 \rho}{3} \frac{(1+\lambda)}{\left(1- \frac{\kappa^2\phi'^2}{6 }\right)} \label{fried2}\,, \\
\!\!\frac{Q_0}{\rho}& =& \frac{\gamma^2 H^2}{2}\Bigg[ \frac{2D}{C}  \phi'' -\frac{2D}{C}  \phi' \!\left(
3\,\omega + \frac{\kappa^2\phi'^2}{2} + \frac{3(1+\omega) B}{2(1+\lambda)} \right) 
+ \left(\frac{D}{C}\right)_{\!\!,\phi} \!\!\phi'^2  + \frac{C_{\!,\phi}}{H^2 C}(\gamma^{-2}-3\omega) \Bigg], \nonumber \\
\eea
where  here we  denote $'= d/dN$. Note also that \eqref{fried2} implies that $\kappa\,\phi' \leq\pm \sqrt{6}$.

Using these equations and further defining a dimensionless scalar field $\varphi = \kappa\,\phi$,  we can rewrite \eqref{friedmann2} and \eqref{kgSimple} as:
\bea
&& H' = - H \left[ \frac{3B}{2(1+\lambda)}   (1+ \omega) + \frac{\varphi'^2}{2}\right], \label{Hprime}\\ \nonumber \\
&&  \varphi'' \left[1\!+\! \frac{3H^2\gamma^2 B}{\kappa^2(1+\lambda)} \frac{D}{C} \right]  
+ 3\,\varphi'  \left[1- \omega \frac{3H^2\gamma^2 B}{\kappa^2(1+\lambda)} \frac{D}{C} \right] +
\frac{H'}{H} \varphi' \left[1+ \frac{3H^2\gamma^2 B}{\kappa^2(1+\lambda)} \frac{D}{C} \right]  
 \nonumber \\
&& \hskip1.5cm  + \frac{3B}{1+\lambda} \alpha(\varphi)(1-3 \,\omega \gamma^2) 
  + \frac{3B\lambda}{(1+\lambda)} \frac{V_{,\varphi}}{V}  + \frac{3 H^2 \gamma^2 B}{\kappa^2(1+\lambda)} \frac{D}{C}\left[ (\delta(\varphi) - \alpha(\varphi)) \,\varphi'^2 \right]   =0\,, 
  \nonumber \\
  \label{phiHeq}
\eea
where we  defined: 
\bea
&&B \equiv 1-\frac{\varphi'^2}{6},  \\
&& \gamma^{-2} = 1- \frac{H^2}{\kappa^2}\frac{D}{C} \varphi'^2\,,  \label{gammaH}\\
&& \alpha(\varphi) = \frac{d \ln C^{1/2}}{d\varphi}, \label{alphaeq}\\
&& \delta(\varphi) = \frac{d \ln D^{1/2}}{d\varphi} \label{deltaeq}\,.
\eea

One can solve the system of coupled equations above for $H$ and $\varphi$ as functions of $N$. However, in some cases it is simpler to use \eqref{Hprime} into \eqref{phiHeq}  and solve the following disformal master equation: 
\bea\label{mastereqsimple}
 \frac{2(1+\lambda)}{3 B} \,\varphi'' &+& \left(2\lambda +1 -\omega \right)\varphi'  + 2 \lambda\, \frac{d\ln  V}{d\varphi}   +2 (1 -3\,\omega\,\gamma^{2})\,\alpha(\varphi)
  \nonumber \\
&+ & \,  \frac{2\gamma^2(1+\lambda)}{3B}\frac{D \rho}{C} \left(\varphi'' -3\,\varphi'
\left[ \omega + \frac{\varphi'^2}{6} +  \frac{(1+\omega) B}{2(1+\lambda)}\right] 
+ \frac{C}{2D}\left(\frac{D}{C}\right)_{\!,\varphi} \varphi'^2 \right) \!=0 \,,  \nonumber \\
\eea
with $\gamma$ given by:
\be
\gamma^{-2} = 1- \frac{(1+\lambda)}{3B}\frac{D \rho}{C} \varphi'^2\,.
\ee

\ni From \eqref{mastereqsimple} we see that the conformal case is recovered for  $D=0$, when the second line vanishes. Moreover, the disformal piece appears always together with derivatives of the scalar field, as expected and  also nontrivially coupled to the energy density. This complicates considerably the analysis of the disformal case, as we will see below.  

\subsection{Modified expansion rate}\label{SecER}

The effect of the expansion rate during the early time evolution due to the  presence of a scalar field can be extracted from the Hubble parameter evolution in the  disformal frame defined as:
$$
\tilde H = d (\log \tilde a )/d\tilde \tau, 
$$ 
which can be written using \eqref{tildea} as:
\be\label{Htilde}
\tilde H = \frac{H\gamma}{C^{1/2}} \left(1+ \alpha(\varphi) \varphi' \right)\,, 
\ee
 where remember that  $\gamma$ depends on $H$  (or $\rho$) as seen from \eqref{gammaH}, while in the pure conformal case  $D=0$ and $\gamma = 1$. Note that in principle, the factor $(1+ \alpha(\varphi) \varphi' )$ can be positive or negative, indicating an expansion or contraction modified rate. We  stick to positive definite values for this factor and therefore only modified expansion rates, though in principle, one could have a brief contraction period during the early universe evolution, before the onset of BBN\footnote{See  \cite{Contraction} for a review on scenarios with a possible contraction phase in the early universe.}. 
 Moreover, notice that while $\tilde H$ can grow during the cosmological evolution, the null energy condition (NEC) is not violated. This is because  the Einstein frame expansion rate $H$ is dictated by the energy density $\rho$ and pressure $p$, which obey the NEC and therefore $\dot H<0$ during the whole evolution, as it should (see for example \cite{Creminelli}).

We further want to relate the modified expansion rate to the expected expansion rate in general relativity (GR), that is: 
\be\label{HGR}
H_{GR}^2 = \frac{\kappa_{GR}^2}{3} \, \tilde \rho\,.
\ee
We can do this be using the Friedmann equation \eqref{fried2} and the relation between the energy densities \eqref{eqrhos} to write
\be\label{Heq2}
\gamma^{-1} H^2 =  \frac{\kappa^2}{\kappa_{GR}^2} \frac{ C^2\,(1+\lambda)}{B} \, H_{GR}^2  \,.
 \ee
Using the definition of $\gamma$ (see \eqref{gammaH}) into this equation, one finds a cubic equation for $H^2$ in terms of all the other parameters. The real positive solution to that equation can then be replaced into  \eqref{Htilde} to find the modified expansion rate $\tilde H$, which will thus be a complicated function of $H_{GR}$ as we now see. 
The cubic equation for $H$ takes the form:
\be\label{eqforH2}
d_1 H^6-H^4 + d_2^2=0\,,
\ee
where 
\bea\label{Eqad}
d_1=\frac{D}{C}\frac{\varphi'^2}{\kappa^2} \,, \hspace{1cm} d_2=
\frac{\kappa^2}{\kappa_{GR}^2}\frac{C^2 (1+\lambda)H_{GR}^2}{B}\,.%
\eea
 The solutions  to \eqref{eqforH2} can be written as
\be\label{Hdis}
H^2=\frac{1}{3 d_1}\left(1+\left(\frac{2}{\Delta}\right)^{1/3}+\left(\frac{\Delta}{2}\right)^{1/3}\right)
\ee

\ni with $\Delta=2-27d_1^2d_2^2+d_1d_2\sqrt{27(27d_1^2d_2^2-4)}$. The other two solutions can be  obtained by replacing 
$$ \left(\frac{2}{\Delta}\right)^{1/3} \to e^{2\pi i/3} \left(\frac{2}{\Delta}\right)^{1/3}
\qquad {\rm and } \qquad \left(\frac{2}{\Delta}\right)^{1/3} \to e^{4\pi i/3} \left(\frac{2}{\Delta}\right)^{1/3}\,.$$
We are interested in real positive solutions for $H^2$. One possibility to get this is to have the imaginary part of $(\Delta/2)^{1/3}$ vanish by requiring  that $\Delta>0$, which is impossible. 
Therefore, the way to obtain real solutions for $H$ is to have  the imaginary parts of $(\Delta/2)^{1/3}$ and  $(\Delta/2)^{-1/3}$ cancel each other, leaving a real positive solution. 

For  this, we need that  $27d_1^2d_2^2 \leq 4$, which implies the following  relation between the conformal and disformal functions: 

\be\label{ConditionD}
\frac{3\sqrt{3} \,D C\, \varphi'^2 (1+\lambda)}{B} \frac{H^2_{GR}}{\kappa_{GR}^2}\leq 2\,.
\ee
Under this condition, we can rewrite $\Delta$ as:
$$\Delta = 2-27d_1^2d_2^2+ i d_1d_2\sqrt{27(4- 27d_1^2d_2^2)}\,, $$
which allows us to define a complex number $Z\equiv \Delta/2$ and it is easy to check that $\bar Z = 2/\Delta$ and thus
 $|Z|^2 =1$. Denoting further $Z_i$  with $i=1,2,3$ denoting the three solutions to $H$ as explained above, the solutions for $H$, \eqref{Hdis} takes the simple form:
 \be\label{Hdis2}
H^2_i=\frac{1}{3 d_1}\left[1+Z_i^{1/3} + \bar Z_i^{1/3}\right] \,,
\ee
and remember that we are interested only in the real positive solution. We can now plug in \eqref{Hdis2}, as well as the expression for $\gamma$ in terms of $H$ into the Jordan frame expansion rate \eqref{Htilde}, can be written as:
\be\label{Htilde2}
\tilde H^2 = \frac{\kappa^2}{\kappa^2_{GR}} \frac{  \gamma^3\,  C(1+\lambda) (1+\alpha(\varphi)\varphi')^2 }{B} \,
\,H_{GR}^2\,, 
\ee
where there is a non-trivial dependence of $H_{GR}$ encoded in 
\be\label{gammai}
\gamma_i = \frac{1}{3d_1d_2}\left[1+Z_i^{1/3} + \bar Z_i^{1/3}\right] \,.
\ee
In the conformal case, $D=0$, $\gamma =1$ and therefore \eqref{Htilde2} is simply
\be\label{ratioH}
\tilde H^2 =  \frac{\kappa^2}{\kappa_{GR}^2}\frac{C (1+\lambda) (1+\alpha(\varphi) \varphi')^2}{B}\, H^2_{GR} \,.
\ee
From  this relation we define a speed-up parameter $\xi$, which will be useful below to measure the departures from the $GR$ expansion rate result:
\be\label{xieq}
\xi \equiv \frac{\tilde H}{H_{GR}}\,. 
\ee

\section{Modifications of the dark matter relic abundances}\label{Sec:2}

In this section we discuss the modifications to the DM relic abundance's  predictions due to modifications of the expansion rate before the onset of nucleosynthesis caused by the presence of a scalar field conformal and disformally coupled to matter. We start  by revisiting the conformal case, discussed originally in \cite{Catena}\footnote{Modifications to the Boltzmann equation due to a conformal coupling in  (non-critical) string theory  have been discussed in \cite{LMN}.}. 
We first solve (numerically) the master equation for the scalar field \eqref{mastereqsimple} in order to compute the modified expansion rate $\tilde H$ and compare it with the standard expansion rate, $H_{GR}$. We  then use this to compute the modifications to the dark matter relic abundances by solving the Boltzmann equation using the modified expansion rate. We start revisiting by the conformal case by exploring a wide range of initial conditions, masses and cross-sections.  We  then  look at an explicit  disformal example. 

Before solving the master equation \eqref{mastereqsimple}, we would like to write it in terms of Jordan  frame quantities $\tilde \omega = \omega \gamma^2$, $\tilde \rho = C^{-2}\gamma^{-1} \rho$. Moreover, the number of e-folds $N$ can be expressed in terms of Jordan frame quantities as follows. In this frame, the entropy is conserved and is given by $\tilde S=\tilde a\, \tilde s$, where $\tilde s= \frac{2\pi}{45} g_s(\tilde T) \tilde T^3$. So, the conservation of entropy and \eqref{tildea} show that $N$ is a function of temperature and the scalar field as:
\be\label{N}
N\equiv\ln\frac{a}{a_0}=\ln\left[\frac{\tilde T_0}{\tilde T}\left(\frac{g_s(\tilde T_0)}{g_s(\tilde T)}\right)^{1/3}\right]+\ln\left[\frac{C_0}{C}\right]^{1/2}.
\ee
Therefore, we can introduce the  parameter, $\tilde N$, defined as
\be\label{Ntilde}
\tilde N \equiv \ln\left[\frac{\tilde T_0}{\tilde T}\left(\frac{g_s(\tilde T_0)}{g_s(\tilde T)}\right)^{1/3}\right].
\ee
and transform to  derivatives w.r.t.~$\tilde N$  (assuming well behaved functions): 
\bea\label{NtotildeN}
\varphi' =  \frac{1}{\left(1-\alpha(\varphi) \frac{d\varphi}{d\tilde N}\right)} \frac{d\varphi}{d\tilde N} \,,\qquad 
\varphi'' = \frac{1}{\left(1-\alpha(\varphi) \frac{d\varphi}{d\tilde N}\right)^3} \left( \frac{d^2\varphi}{d\tilde N^2} + \frac{d\alpha}{d\varphi}  \left(\frac{d\varphi}{d\tilde N}\right)^3\right)\,. \\ \nonumber 
\eea
In a slight abuse of notation and to keep expressions  neat, in what follows we  denote derivatives w.r.t.~$\tilde N$ with a prime $'$.

\subsection{Conformal case}

We start  with the pure conformal case. That is, we take  $D(\phi) =0$ in \eqref{mastereqsimple} and therefore $\gamma=1$ (and  $\tilde \omega = \omega$). Moreover, during the radiation and matter dominated eras, of interest for us,  the potential energy  of the scalar field is subdominant and therefore, we take  $\lambda \sim 0$. 
Therefore the  master equation \eqref{mastereqsimple} simplifies to:
\bea\label{masterConf}
 \frac{2}{3 (1-\varphi'^2/6)} \,\varphi'' &+& \left(1 -\tilde \omega \right)\varphi'  + 2 (1 -3\,\tilde \omega)\,\alpha(\varphi) =0, 
\eea
which in terms of derivatives wrt $\tilde N$ takes the form:
\bea\label{masterConf2}
\frac{1 }{3B\left[1-\alpha(\varphi) \varphi' \right]^3} \left(\varphi'' + \frac{d\alpha}{d\varphi}  \left(\varphi'\right)^3\right)
+ \frac{(1-\tilde \omega)  }{\left[1-\alpha(\varphi)\varphi'\right]} \,\varphi'
+(1-3\,\tilde \omega) \, \alpha(\varphi)=0 \,, \nonumber \\
\eea
where  $B= 1-\frac{\left(\varphi'\right)^2 }{6\left(1-\alpha(\varphi) \varphi'\right)^2} $. 
Using the relation between $\tilde H$ and $H_{GR}$ defined in \eqref{ratioH}, we can write the speed-up parameter as 
\be
\xi = \frac{\tilde H}{H_{GR}} = \frac{C^{1/2}(\varphi)}{C^{1/2}(\varphi_0)} \frac{1}{\left(1-\alpha(\varphi) \varphi'\right)\sqrt{B}} \frac{1}{\sqrt{1+ \alpha^2(\varphi_0)}}
\ee
where we have used the relation between the bare gravitational constant and that measured by local experiments for  conformally coupled theories \cite{Nordtvedt:1970uv}:  
\be\label{ratioG}
\kappa_{GR}^2 = \kappa^2 C(\varphi_0) [1+\alpha^2(\varphi_0)]\,,
\ee
where $\varphi_0$ is the value of the scalar field at present time.

\subsubsection{Expansion rate modification}\label{ERM}

The scalar equation in the conformal case \eqref{masterConf}, as function of $N$ (for  $\lambda =0$)  contains a term which can be interpreted as an effective potential, dictated by $V_{eff} = \ln C^{1/2}$. For a strictly radiation dominated era, $\tilde \omega =1/3$, the effective potential term vanishes and we are left with an equation that can be solved analytically \cite{Coc}, giving  $\varphi' \propto e^{-N}$. That is, any initial velocity will rapidly go to zero (remember that from the Friedmann equation \eqref{fried2}, $\varphi'$ is constrained to be $\varphi' \lesssim \pm \sqrt{6}$).  
Therefore we explore the effects of having a non-zero initial velocity in our analysis below (see also Appendix \ref{NumImp} for further examples). 
Since the scalar field is expressed in Planck units, we focus on order one or smaller field variations $\Delta\varphi$. One can check, using the analytic solution to \eqref{masterConf} deep in the radiation era, that for initial velocities $\varphi'_0 \ll \pm \sqrt{6}$, the total field displacement is of order $\Delta \varphi \sim \varphi'_0$ \cite{Coc}.  However, given that we don't know much about the theory before BBN, we  explore different initial values for $(\varphi_0, \varphi'_0)$ and study their 
consequences. In particular we explore initial values $\varphi_0$ and $\varphi'_0 \in (-1.0, 1.0)$.
 
We now concentrate on an explicit  conformal factor. We  use  the same conformal factor as that studied in \cite{Catena}, which is given by:
\be\label{Cfactor}
C(\varphi) = (1+ b \,e^{-\beta\, \varphi})^2
\ee
with  the values $b=0.1$, $\beta=8$, which have been shown to satisfy the constraints imposed by tests of gravity, for the parameters $\alpha, \beta$, $\xi$. As we will see, the requirement of reaching the  GR expansion  rate value  by the time of the onset of BBN, drives these parameters to very small values, which are thus consistent with the constraints from gravity for their values today.

As we  discussed above, in the equation of motion for $\varphi$, with $\tilde\omega\neq 1/3 $, the conformal factor acts as an effective potential 
on which the scalar field moves, damped by the Hubble friction (see \eqref{kgSimple}). 
Since any initial velocity $\varphi'$ goes rapidly to zero deep in the radiation era, in the subsequent evolution of $\varphi$,  the term $(\varphi' )^2$  in the master equation will be negligible. In this regime, the equation is that of a particle moving in an effective potential with a damping term. Therefore, one can understand the evolution of $\varphi$  from the form of the effective potential ($V_{eff}=\ln C^{1/2}$) and the initial conditions chosen. For the conformal function we are considering \eqref{Cfactor}, one sees that there is a set of initial conditions that will give rise to an interesting behaviour in $\varphi$, and therefore an interesting modified expansion rate in the Jordan frame $\tilde H$ \eqref{ratioH}, as we now explain.

\begin{figure}[h!]
\centerline{
\includegraphics[width=.65\textwidth]{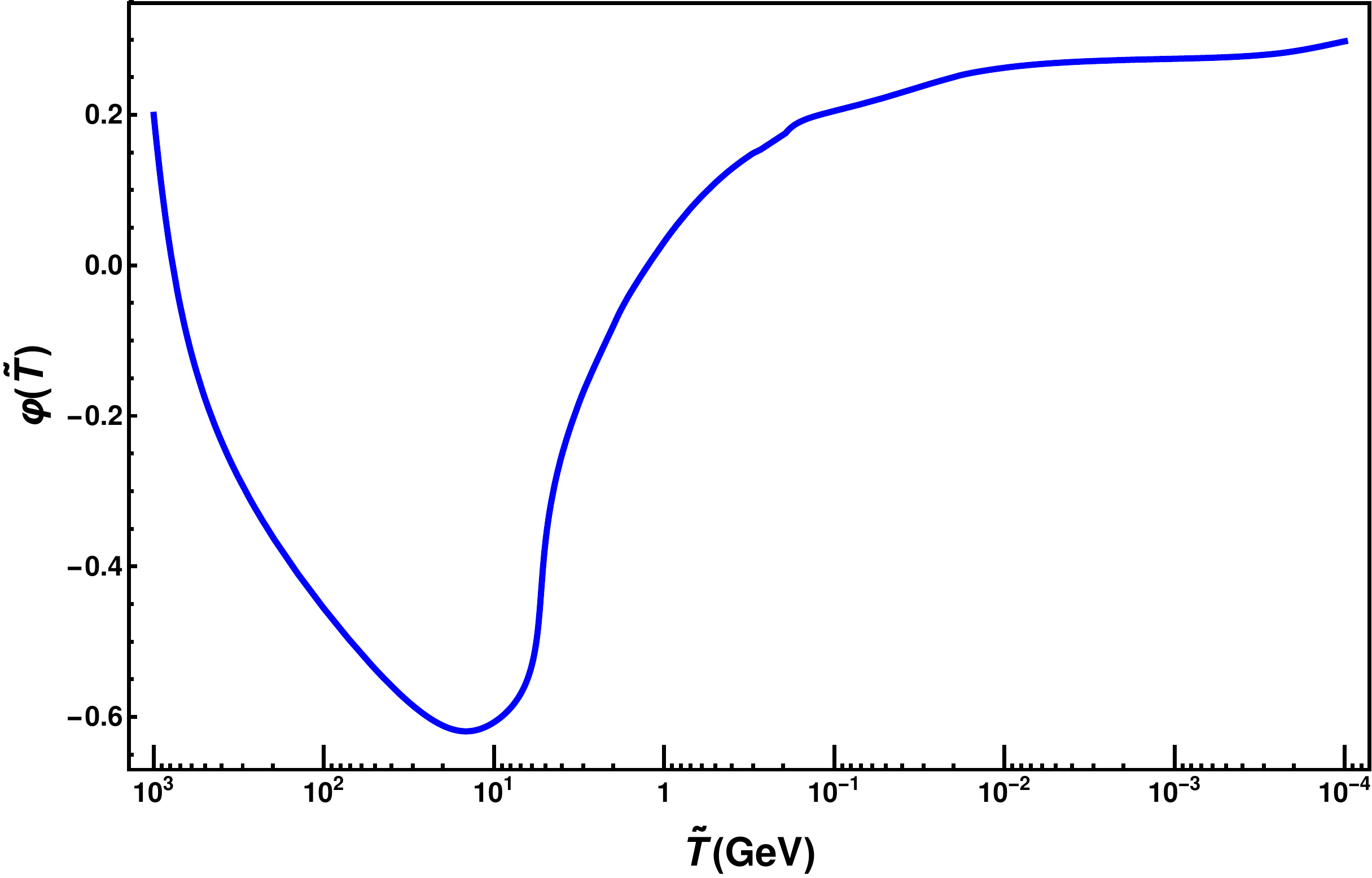}}
\caption{Typical evolution of the scalar field  as temperature decreases. The initial values are $(\varphi,d\,\varphi/d\,\tilde N) = (0.2,-0.994)$.}
\label{FieldFig}
\end{figure}

\begin{figure}[h!]
\centerline{
\includegraphics[width=.65\textwidth]{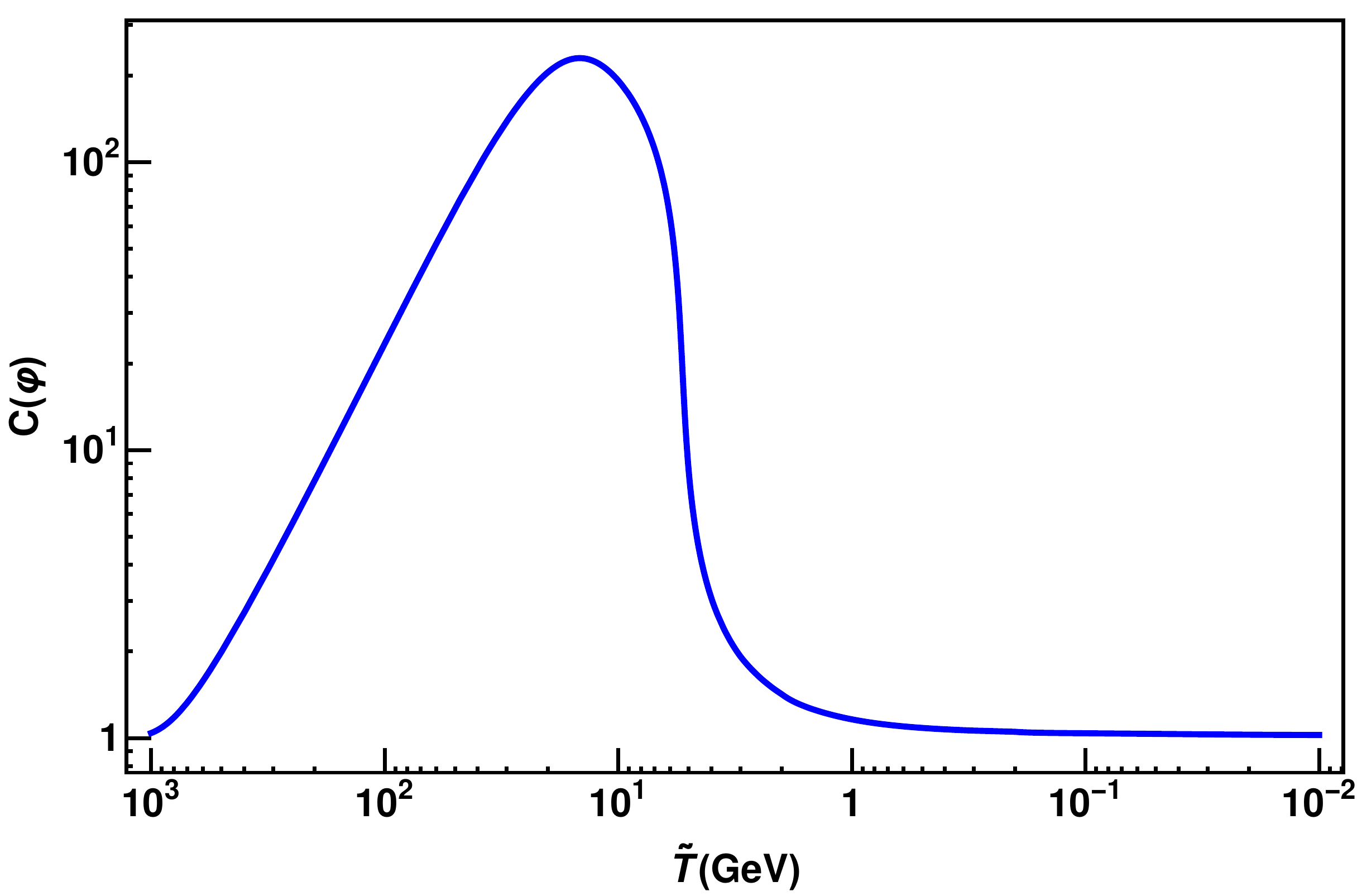}}
\caption{Behaviour of the conformal factor, $C(\varphi)$ as a function of the temperature for the same initial values as in Fig.~\ref{FieldFig}.}
\label{CFig}
\end{figure}

In general, both the initial position and velocity of the scalar field can take any value, positive and negative. In the runaway effective potential dictated  by $\ln C^{1/2}$ for the conformal factor \eqref{Cfactor} we consider here, we have the following possibilities. i) The scalar field starting somewhere up in the runaway effective potential with zero initial velocity. In this case the scalar field will  roll-down the potential, eventually stopping due to Hubble friction, at some constant value of $\varphi$, which depends on the its initial value. So long as $(1+\alpha(\varphi) \,\varphi')$  stays positive (see \eqref{Htilde}),  $C$  will evolve rapidly towards $1$.
 More generally, the initial velocity can be different from zero. If the initial velocity is positive, the behaviour will be similar to the previous case. The field will roll-down the effective potential towards its final terminal  value.

ii) A more interesting possibility arises when one allows for negative initial velocities. In this case, the  field will start rolling-up the effective potential towards smaller values of the field, eventually turning back down and moving  towards its terminal value. 
It is easy to see that an interesting effect happens when the field starts at an initial positive value. Given a sufficient initial negative velocity the field will move towards negative values until its velocity becomes zero and then positive again, as it rolls back down the effective potential. This change in sing for the scalar evolution will  produce a pick in the conformal function that will give rise to a non-trivial modification of the Jordan's frame expansion rate $\tilde H$, as we are looking for. 
As mentioned before, we are interested in (sub-)Planckian initial values $(\varphi_0, \varphi'_0)$, such that $\tilde H>0$. With these requirements, one can  see that given an initial negative velocity, there is a suitable initial value of the scalar field such that the behaviour just described holds and the expansion rate $\tilde H$ has an interesting evolution before the onset of BBN. At late times the conformal function goes to one and the GR expansion is recovered. 
 We show this behaviour explicitly in Figures \ref{FieldFig} and \ref{CFig} where we plot the numerical solution for the evolution of $\varphi$ and  $C(\varphi)$  as functions of the temperature. In these plots we find $\varphi=\varphi(\tilde T)$ by first solving \eqref{masterConf2} numerically with initial conditions $ (\varphi_0,\varphi'_0) = (0.2,  -0.99)$  and then use \eqref{Ntilde} to express $\varphi(\tilde N)$ as function of $\tilde T$ \footnote{In appendix \ref{NumImp} we show further examples of the thermal  evolution of the scalar field.}.
 As we can see,  the conformal factor  starts growing  towards a maximum value as $\varphi$ moves to negative values, to rapidly drop down towards its GR value at $C\to 1$ as $\varphi$ moves down the effective potential towards positive values. This non-trivial effect will give rise to the possibility of re-annihilation, as we discuss below.
 

\begin{figure}[h!]
\centerline{
\includegraphics[width=.8\textwidth]{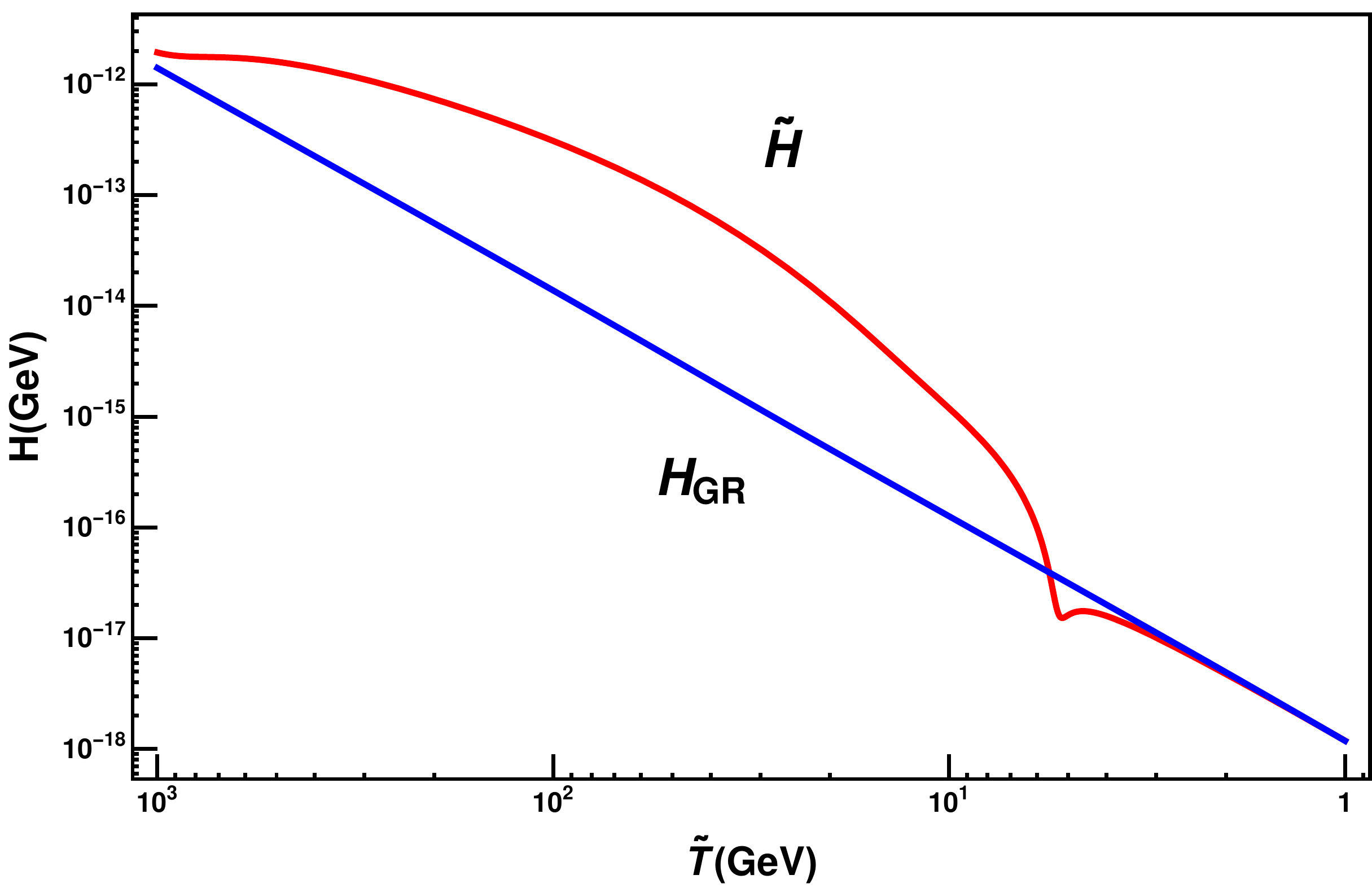}}
\caption{Comparing the Hubble expansion rate $\tilde H$ in the Jordan Frame with the standard Hubble expansion rate $H_{GR}$. 
The presence of the scalar field enhances and decreases the expansion rate during the radiation dominated era. This plot corresponds to
initial conditions given by $(\varphi_0,\varphi'_0) = (0.2,-0.994)$.}
\label{HubblesFig}
\end{figure}

Based on the discussion above, we have solved the master equation \eqref{masterConf2}, to find the  the scalar field  as a function of $\tilde N$ for various initial conditions,
where we see the interesting behaviour explained above. The resulting modified expansion rate and its comparison with the standard case is  shown in Figure \ref{HubblesFig} for 
the same initial conditions as in Figures \ref{FieldFig} and \ref{CFig}. In our numerical exploration, we choose initial conditions for which the notch in the expansion rate
(see Fig.~\ref{HubblesFig}) occurs closer to the BBN time\footnote{In appendix \ref{NumImp} we show more examples
of modified expansion rate using different initial conditions.}. This has interesting consequences for the dark matter annihilation, as we discuss below.


%
When solving the master equation \eqref{masterConf2}, we have taken into account an important effect that occurs during the radiation dominated era. Deep in this epoch, the equation of state is   given by $\tilde\omega=1/3$. When a particle species in the cosmic soup becomes non-relativistic, $\tilde\omega$ differs slightly from $1/3$. 
When the temperature of the universe drops below the rest mass of each of the particle types, there are non-zero contributions to $1-3\tilde \omega$. This activates the 
effective potential, which can be seen in the last term of \eqref{masterConf2}, and  displaces, or ``kicks`` the field along $V_{eff}$.

To examine this effect in more detail, we start by writing $1-3\,\tilde \omega$ during the early stages of the universe as in \cite{Catena} and \cite{MW}
\be\label{eqomega}
1-3\,\tilde\omega = \frac{\tilde \rho - 3\, \tilde p}{\tilde \rho} =\sum_{A} \frac{\tilde \rho_A - 3\tilde p_A}{\tilde \rho} +\frac{\tilde \rho_m}{\tilde \rho}\,,
\ee
where the sum runs over all particles in thermal equilibrium during the radiation dominated era and $\tilde \rho_m$ is the contribution
from the non-relativistic decoupled and pressureless matter. The summation over all the particle is responsible for the kicking effect discussed above. Then, a kick 
function is defined as
\be\label{kick}
\Sigma(\tilde T) \equiv \sum_{A} \frac{\tilde \rho_A - 3\tilde p_A}{\tilde \rho}\,,
\ee
where the energy density $\tilde\rho_A$ and pressure $\tilde p_A$ of each type $A$ of particle are given by
\be\label{erhoA}
\tilde \rho_A(\tilde T) =\frac{g_A}{2\pi^2}\int^\infty_{m_A}\frac{\left(E^2-m_A^2\right)^{1/2}}{\exp(E/\tilde T)\pm 1}E^2dE
\ee
\be\label{epA}
\tilde p_A(\tilde T) =\frac{g_A}{6\pi^2}\int^\infty_{m_A}\frac{\left(E^2-m_A^2\right)^{3/2}}{\exp(E/\tilde T)\pm 1}E^2dE
\ee
with $g_A$ being the number of internal degrees of freedom of species of type $A$ and the plus (minus) sign in the integral corresponds to fermions (bosons).
To compute \eqref{kick}, we use the Standard Model particle spectrum. In particular, we take into account the top quark, the Higgs boson, Z boson, W bosons, 
bottom quark, tau lepton, charm quark, charged pions, neutral pion, moun lepton and the electron\footnote{We present more details of the kick function in appendix \ref{NumImp}.}. As we show in Figure \ref{sigmafig}, 
$\Sigma$ is mostly zero, except when the kicks happen.

During the radiation dominated era $\tilde \rho \simeq \pi^2 g_{eff}(\tilde T) \tilde T^4/30$, where $\tilde T$ is the Jordan frame temperature and
$g_{eff}$ is the total number of relativistic degrees of freedom. Also, during this stage $\tilde \rho_m$ is negligible and as we have shown $\Sigma$ is slightly different 
than zero. Therefore, we can compute the equation of state from  \eqref{eqomega}   as  $\tilde\omega=(1-\Sigma(\tilde T))/3$.

\begin{figure}[h!]
\centerline{
\includegraphics[width=.65\textwidth]{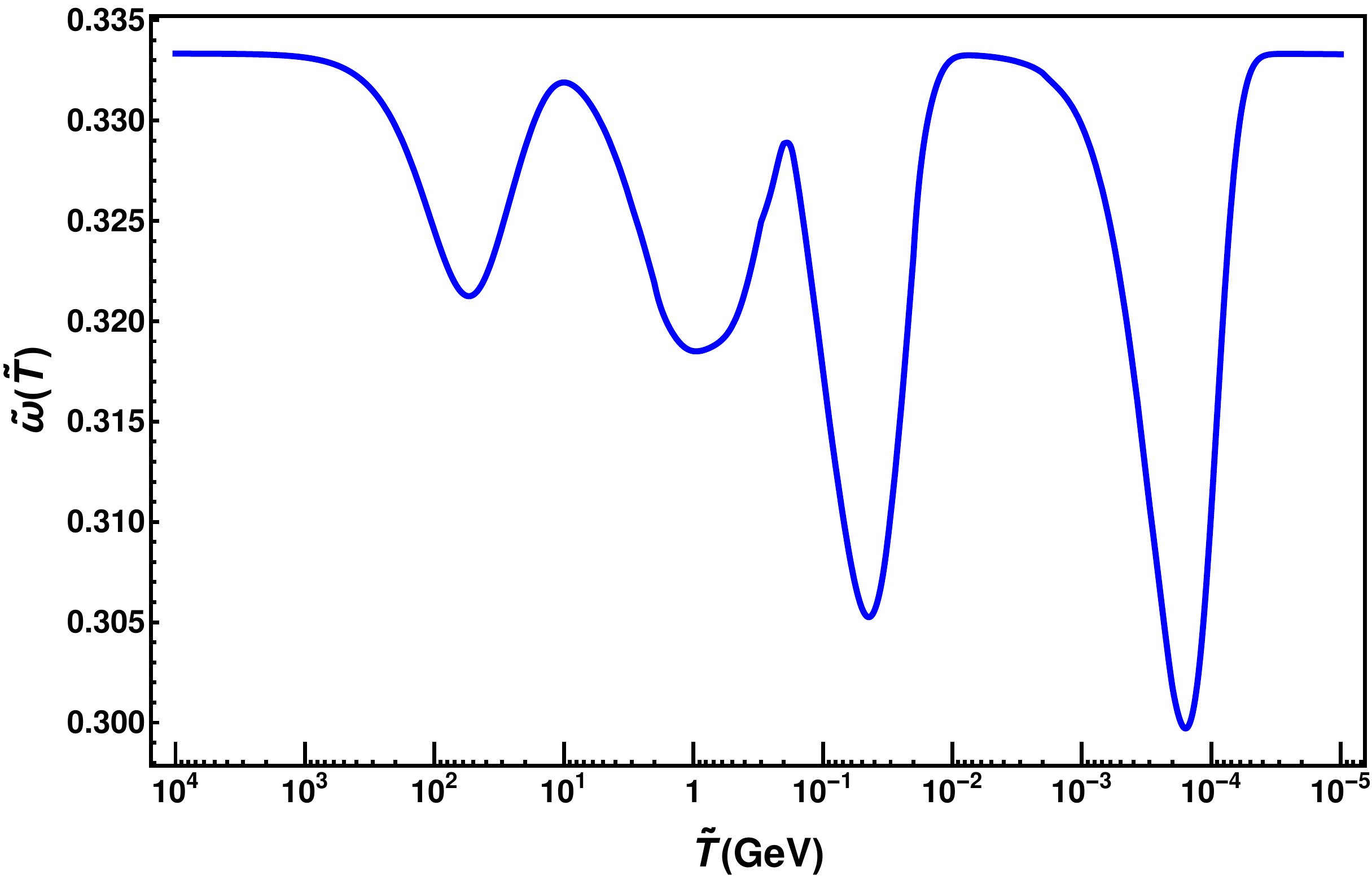}}
\caption{Evolution of $\tilde\omega$ in \eqref{eqomega} as function of temperature during the radiation dominated era.}
\label{plotomegaradiation}
\end{figure}

In Figure \ref{plotomegaradiation} we show the evolution of $\tilde \omega$ between 10 TeV and 10 eV. This figure shows four troughs, which are the ``kicks'' mentioned above. 
Each kick corresponds to the transition of one or more particles to the non-relativistic regime. For example, the trough at around 0.5 MeV is due to the electron, while the one
at around 100 GeV is due to the heavy particles ($t$, $H$, $Z$ and $W$).

Towards the end of the radiation era, approaching the transition to the matter dominated era, eq.~\eqref{eqomega} takes the approximate form:
\be\label{lateomega}
 1-3\,\tilde\omega \simeq \frac{\tilde \rho_{m}}{\tilde \rho_m + \tilde \rho_r} \simeq \frac{1}{1+ \tilde T/\tilde T_{eq}} \,,
\ee
where $\tilde T_{eq} \sim {\cal O}(10^{-9})$GeV is the temperature at matter-radiation equality, that is, $\tilde \rho_m(\tilde T_{eq}) = \tilde \rho_r(\tilde T_{eq})$. 
We can now combine \eqref{kick} and  \eqref{lateomega} to compute the thermal evolution of \eqref{eqomega} in the radiation dominated and matter dominated
eras and use it in the master equation.
%

%

\subsubsection{Parameter Constraints}\label{ParCons}

In scalar-tensor theories of gravity, there are some constraints on the parameters that need to be taken into account. 
Deviations from GR can be parametrised in terms of the post-Newtonian parameters $\gamma_{PN}$ and $\beta_{PN}$, which are given in terms of  $\alpha(\varphi_0)$ defined in \eqref{alphaeq} and  its derivative $\alpha'_0= d\alpha/ d\varphi |_{\varphi_0}$  as  \cite{Esposito,Will}:
\be
\gamma_{PN} -1 = -\frac{2\alpha_0^2}{1+\alpha_0^2} \,, \qquad \beta_{PN} -1 = \frac12\frac{\alpha'_0\alpha_0^2}{(1+\alpha_0^2)^2} \,,
\ee
Solar system  tests of gravity, including the perihelion shift of Mercury, Lunar Laser Ranging experiments, and the measurements of the Shapiro time delay   by the Cassini spacecraft \cite{Lunar,Shapiro,Cassini} indicate that $\alpha_0$ should be very small, with values $\alpha_0^2 \lesssim 10^{-5}$, while binary pulsar observations impose that $\alpha'_0\gtrsim -4.5$.
The last constraint applies to the the speed-up factor $\xi$, which has to be of order 1 before the onset of BBN. In our  examples we have 
$\alpha_0^2 \simeq 2\times10^{-5}$, $\alpha'_0 > 0$ and $\xi\approx1.05$.

\subsubsection{Impact on relic abundances }\label{Impact}

\begin{figure}[h!]
\centerline{
\includegraphics[width=.85\textwidth]{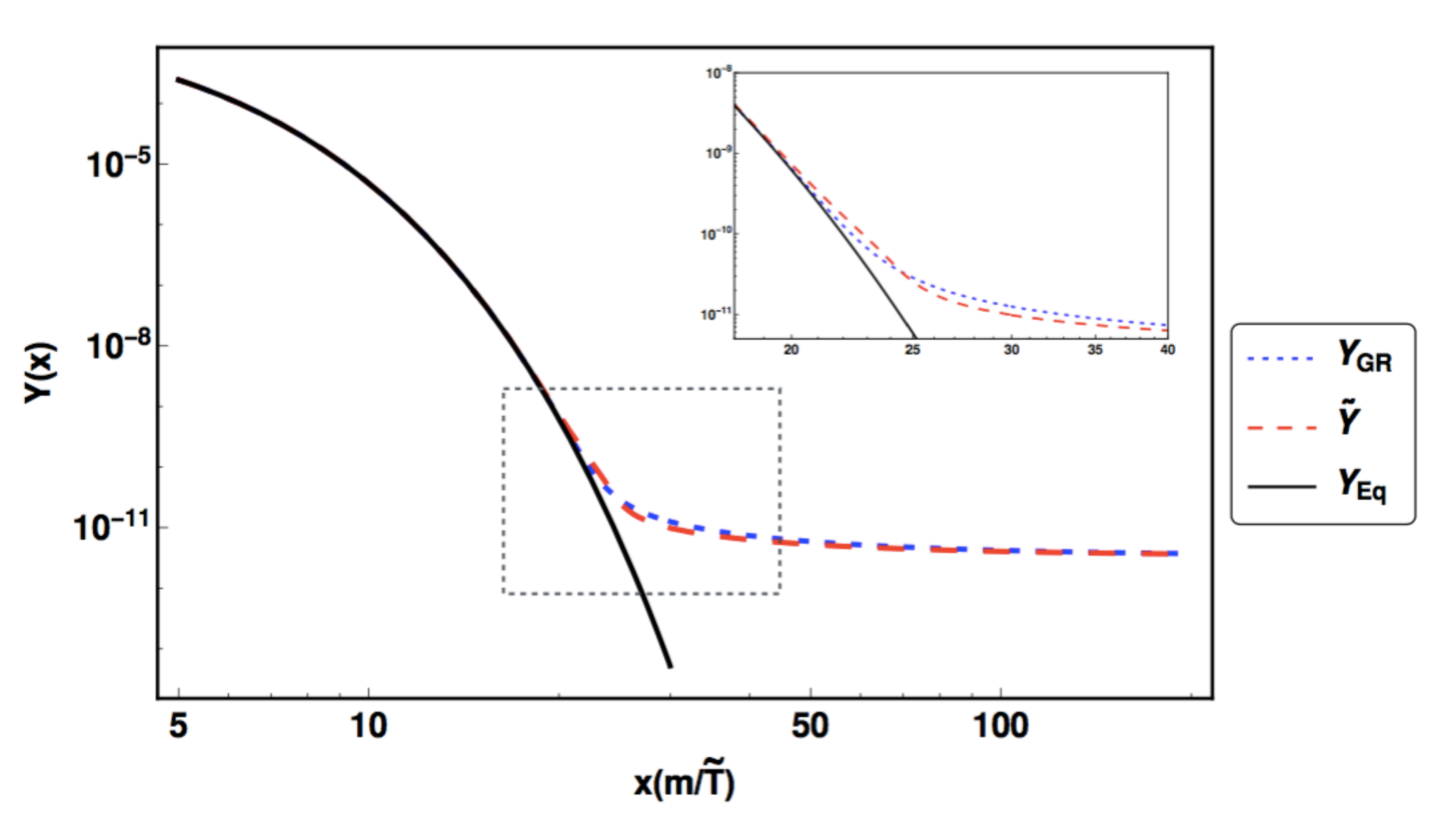}}
\caption{Evolution of the abundance as temperature changes for a DM particle of mass 130 GeV.}
\label{Y130Fig}
\end{figure}

\begin{figure}[h!]
\centerline{
\includegraphics[width=.65\textwidth]{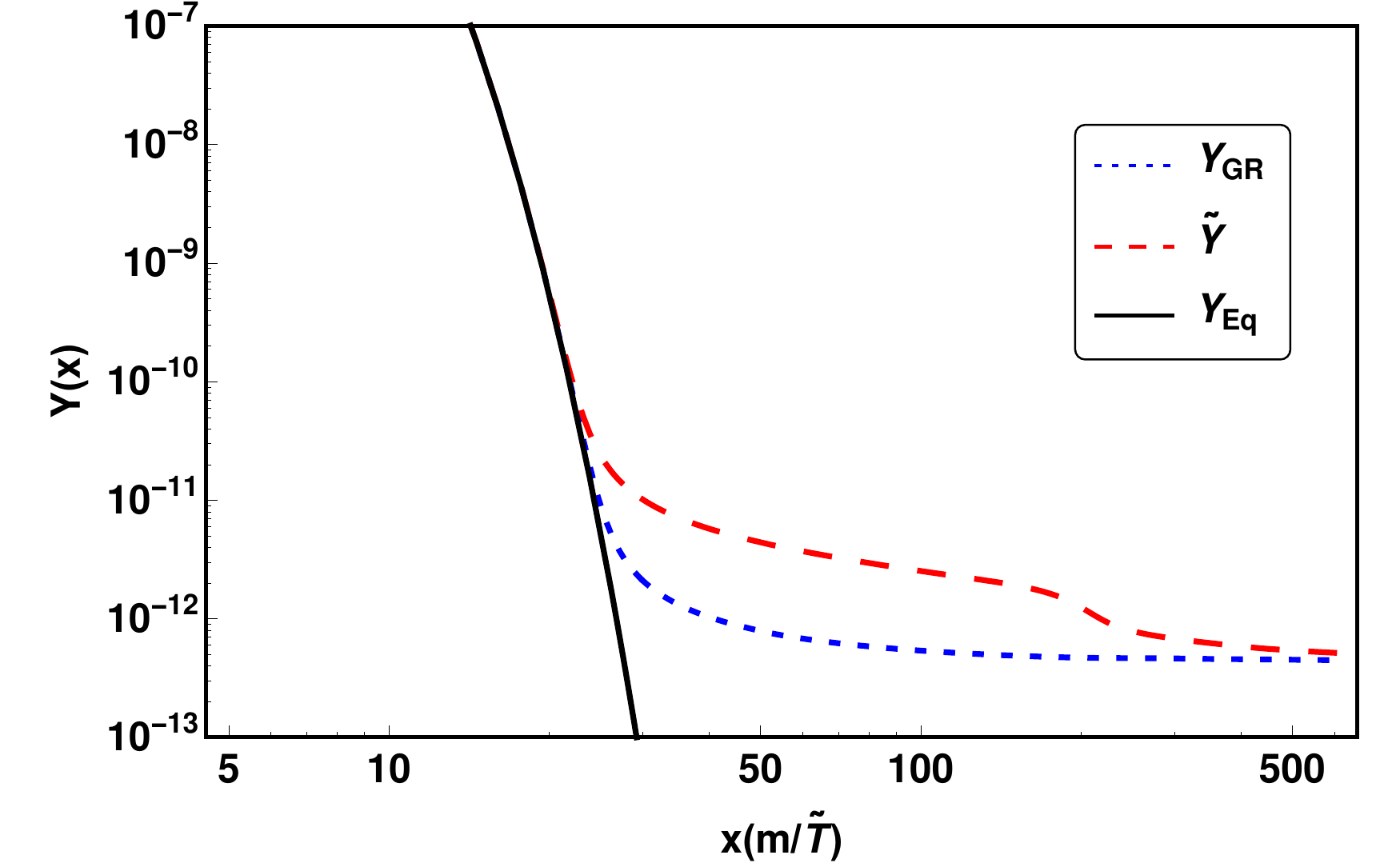}}
\caption{Abundance for a mass of 1000 GeV.}
\label{Y1000Fig}
\end{figure}

We are now ready to discuss the impact of the modified expansion rates on the relic abundance of dark matter species. 
For a dark matter species $\chi$ with mass $m_\chi$ and annihilation cross-section $\langle \sigma v \rangle$, where $v$ is the relative velocity, the dark matter number density $n_\chi$ evolves according to the Boltzmann equation 
\be\label{Boltzn}
\frac{d n_\chi}{dt} = -3 \tilde H n_\chi - \langle \sigma v \rangle \left( n_\chi^2  - (n_\chi^{eq})^2 \right)\,,
\ee
where, as we have discussed above, the relevant  expansion rate  is the Jordan frame one, which can give interesting effects due to the presence of the scalar field. Further $n_\chi^{eq}$ is the equilibrium number density. We can  rewrite this equation in terms of $x=m_\chi/\tilde T$
\be\label{Boltzy}
\frac{d Y}{dx} = - \frac{\tilde s \langle \sigma v \rangle }{x \tilde H}  \left( Y^2  - Y_{eq}^2 \right) \,. 
\ee
where $Y = \frac{n_\chi}{\tilde s}$, $\tilde s= \frac{2\pi}{45} g_s(\tilde T) \tilde T^3$.
Numerical solutions to the Boltzmann equation \eqref{Boltzy} with the modified expansion rate $\tilde H$ were found for dark matter particles
with masses ranging from 5 GeV to 1000 GeV. For instance, we show solutions in figures \ref{Y130Fig} and \ref{Y1000Fig} 
for two different masses. As we can see from \eqref{Boltzy}, the annihilation cross-section influences the evolution of the abundance $Y$. The current
value of $Y$ determines the present dark matter  content of the universe. This can be seen clearly by recalling the current value of the energy density parameter
$\Omega_0=\frac{\rho_0}{\rho_{c,0}}=\frac{m\,Y_0\,s_0}{\rho_{c,0}}$, where $\rho_{c,0}$ and $s_0$ are the well-known current values of the critical energy
density and the entropy density of the universe, respectively. So, for each single mass, the thermally-averaged annihilation
cross section, $\langle\sigma v \rangle$, was chosen such as the current DM content of the universe is 27 \%, so $\Omega_0=0.27$. 

\begin{figure}[h!]
\centerline{
\includegraphics[width=.65\textwidth]{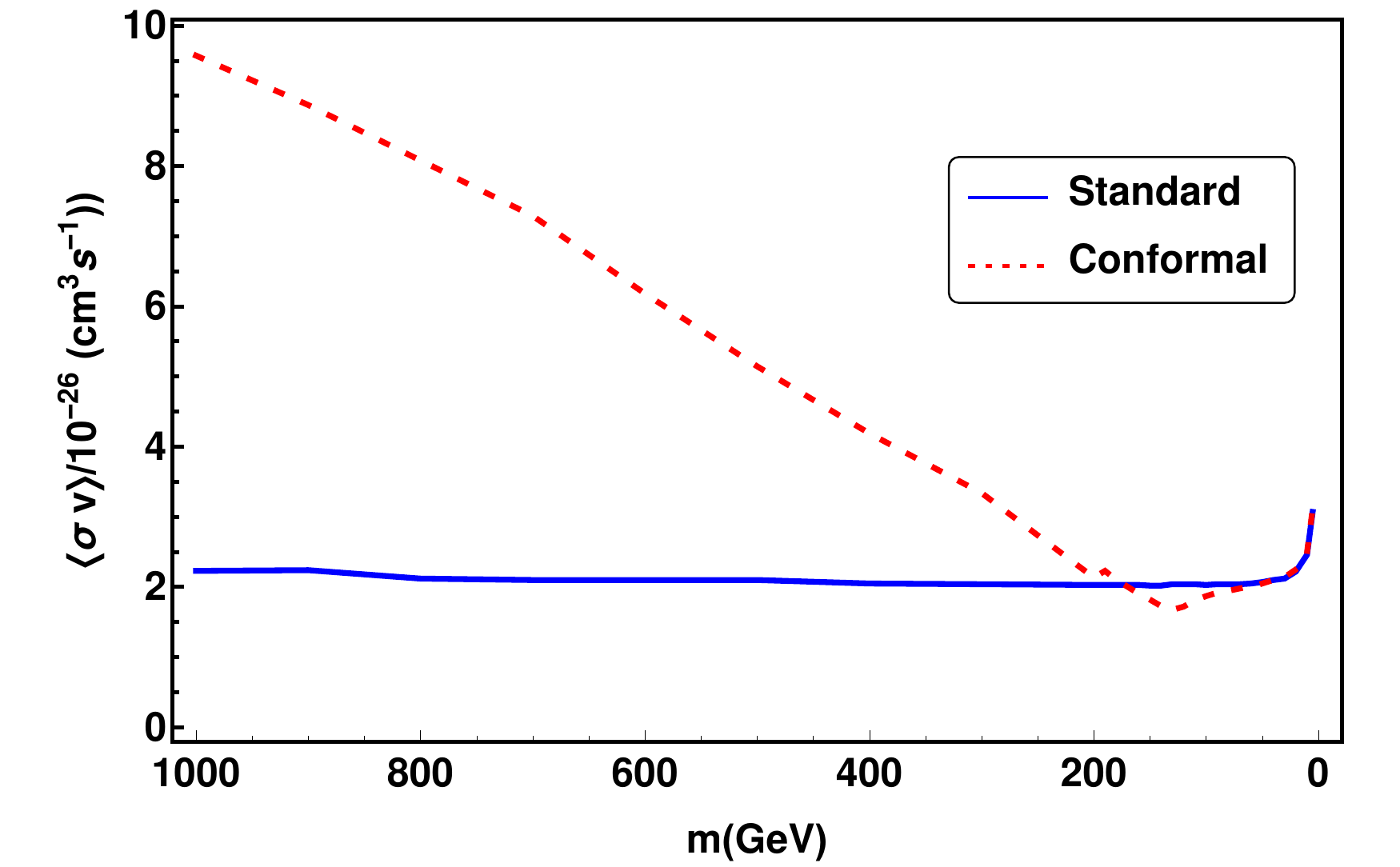}}
\caption{Annihilation cross section as function of mass. The presence of the scalar field enhances the $\langle\sigma v \rangle$ for large masses,
and diminishes $\langle\sigma v \rangle$ for masses around 130 GeV, while small mass the effect is almost negligible. }
\label{AnniFig}
\end{figure}

In Figure \ref{AnniFig} we show the annihilation cross-section, 
$\langle\sigma v \rangle_{Conformal}$, found for all masses and compare it to the annihilation cross sections for the standard cosmology model, 
$\langle\sigma v \rangle_{Standard}$. As it is shown, for large masses $\langle\sigma v \rangle_{Conformal}$ is larger than $\langle\sigma v \rangle_{Standard}$,
up to a factor of four. As the mass decreases  $\langle\sigma v \rangle_{Conformal}$ decreases up to the point where is smaller than $\langle\sigma v \rangle_{Standard}$. 
Then, for masses smaller than 100 GeV the figure shows that $\langle\sigma v \rangle_{Conformal} \approx \langle\sigma v\rangle_{Standard}$. 
Thus, we have found that the annihilation cross-sections can be larger or smaller than the thermal average cross-section.
Just to give an example of larger and smaller cross-section, the following table compares the numerical values of $\langle\sigma v \rangle_{Conformal}$ and $\langle\sigma v \rangle_{Standard}$ for two dark matter masses, 1000 GeV and 130 GeV.

\begin{center}
 \begin{tabular}{|c|c|c|}
 \hline
  Mass (GeV) & $\langle\sigma v \rangle_{Conformal} (\times 10^{-26}\rm{cm^3/s})$&$\langle\sigma v \rangle_{Standard} (\times 10^{-26}\rm{cm^3/s})$\\ \hline
   1000 &9.57 &2.23\\ \hline
    130 &1.68 &2.04\\ \hline
 \end{tabular}
\end{center}

Figures \ref{Y130Fig} and \ref{Y1000Fig} show the evolution of the abundance $\tilde Y(x)$
for DM particles with masses 130 GeV and 1000 GeV, respectively. These figures also include the abundance $Y_{GR}(x)$ calculated in the standard 
cosmology model and the equilibrium abundance $Y_{Eq}(x)$.

The temperature evolution of the abundance for a 130 GeV mass is not noticeable affected by the presence of the scalar field $\phi$. In this case, $\tilde Y$
and $Y_{GR}$ are almost indistinguishable from one another.
On the other hand, the scalar field $\phi$ has a prominent effect on the temperature evolution of the abundance for a 1000 GeV DM particle. First of all,
the freeze-out happens earlier than expected due to the enhancement of the expansion rate, $\tilde H$. Then, an unusual effect appears. As the temperature decreases, $\tilde H$ becomes smaller than the interaction rate\footnote{The interaction rate is defined as $\tilde\Gamma\equiv\langle\sigma v \rangle_{Conformal}\,\tilde s\,\tilde Y$.} 
 $\tilde \Gamma$ and a short period of annihilation starts again called ``re-annihilation''. The re-annihilation process reduces the abundance of dark matter until a second and final freeze-out happens.  After this final freeze-out the abundance remains constant.


\begin{figure}[h!]
\centerline{
\includegraphics[width=.85\textwidth]{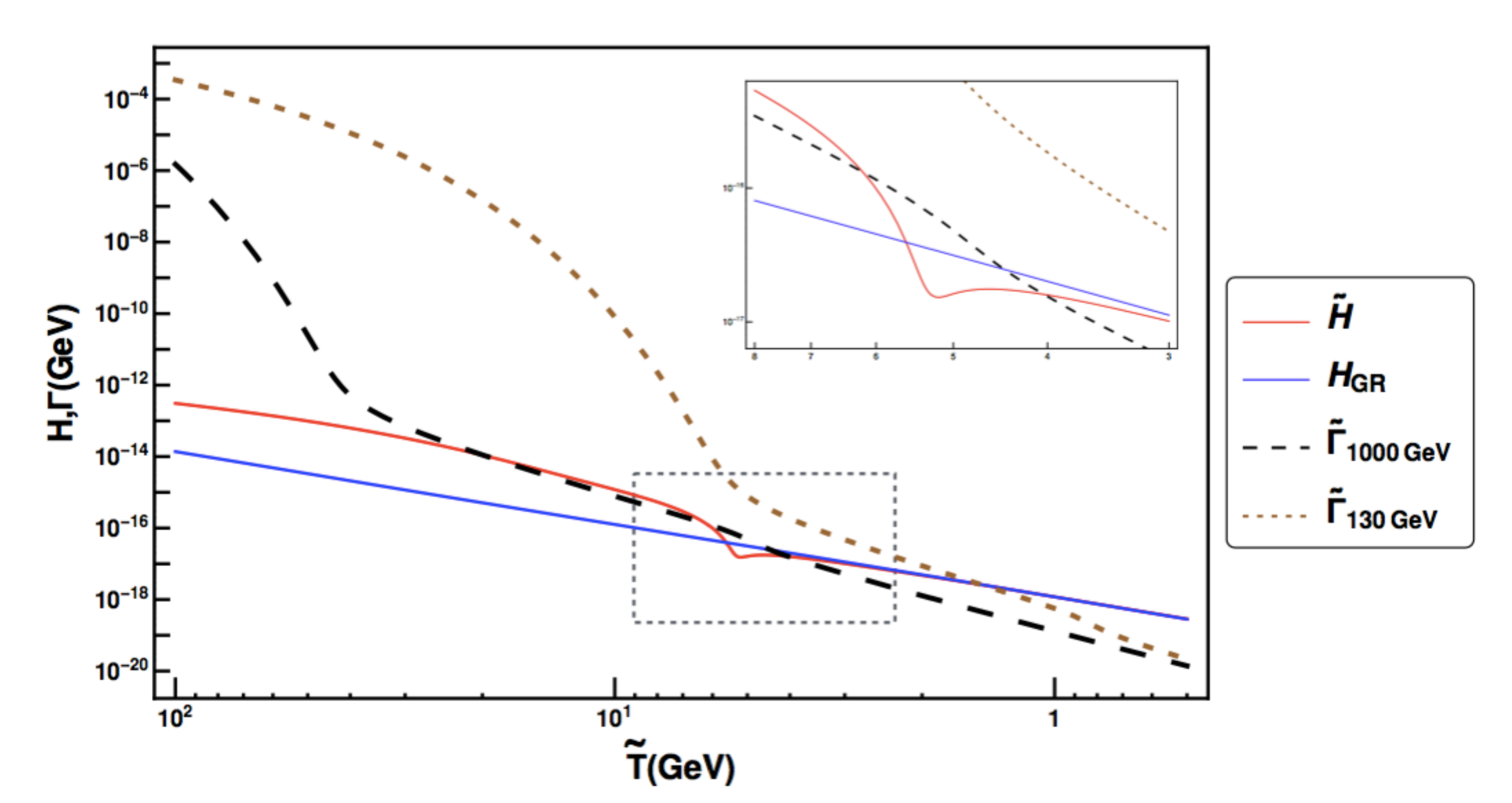}}
\caption{Expansion rate (as in figure \ref{HubblesFig}) and interaction rate as function of temperature. The interaction rate, $\tilde \Gamma$, is given by $\langle\sigma v \rangle_{Conformal}\,\tilde s\,\tilde Y$. We use $\tilde Y$ from figures \ref{Y130Fig} and \ref{Y1000Fig} and the values of $\langle\sigma v \rangle_{Conformal}$ presented previously for 130 GeV and 1000 GeV masses.  }
\label{plotsHgs} 
\end{figure}
The re-annihilation phase can be described better by discussing the relation between the expansion rate $\tilde H$  and the interaction rate $\tilde\Gamma$.
The first freeze-out happens when $\tilde\Gamma$ becomes smaller than $\tilde H$  which can be seen in figure \ref{plotsHgs} to happen around a temperature of 50 GeV
for a 1000 GeV particle. Then, near to 7 GeV $\tilde H$  drops below $\tilde\Gamma$, and so the re-annihilation process starts and goes on until the second freeze-out occurs. Around 2 GeV $\tilde H$ becomes much larger than $\tilde\Gamma$ and so the abundance becomes almost constant. 
Our analysis shows that, as found in \cite{Catena}, re-annihilation occurs for this particular choice of conformal factor. However, we found that when fully integrating the master equation, the re-annihilation occurs only for very large masses of the dark matter particles (in \cite{Catena} it was found for $m=50$GeV). 
On the other hand, in \cite{MW}, no re-annihilation was found\footnote{Although \cite{MW} used a different conformal factor to \cite{Catena},  we expect the re-annihilation effect to be present also in that case.}, which was probably due to the initial conditions used and the values of the DM masses explored.

\subsection{Disformal case}

We  now discuss briefly the effect of the disformal factor in the metric \eqref{gtilde} to the expansion rate of the universe, $\tilde H$, and compare it to the conformal modification to $\tilde H$\footnote{We leave a detailed exploration for a future publication.}. Hence, we explore $D(\phi)\neq0$ for the same
conformal factor studied before, that is, $C(\varphi) = (1+ b \,e^{-\beta\, \varphi})^2$ for $b=0.1$, $\beta =8$. To investigate these modifications, we first need to look at the the scalar field evolution with temperature.

In the pure conformal case studied above, we found the thermal evolution of the scalar field by solving the master equation \eqref{masterConf2} numerically, which is  \eqref{mastereqsimple} for 
$D(\phi)=0$.  However, to study the effects of the disformal factor on the scalar field, 
it is more convenient to solve the system of two coupled equations \eqref{Hprime} and
\eqref{phiHeq}. Using these equations we find solutions for the  dimensionless scalar field $\varphi$,  and  for the expansion rate in the Einstein frame $H$.

Notice that   solving the system of coupled equations  or solving 
the master equation to find the thermal evolution of the scalar field are equivalent methods (as we have explicitly checked), because \eqref{mastereqsimple} it is nothing but a 
combination \eqref{Hprime} and \eqref{phiHeq}. However, while in the pure conformal case the master equation can be made independent of $H$ (or $\rho$), this is not the case for the more general disformal case, as we can see in eq.~\eqref{mastereqsimple}.

In the same way as for the conformal case, we are interested mainly in the radiation and matter eras and therefore we can neglect the potential energy of the scalar field. Thus, we consider $V\sim 0$ and $\lambda=0$. Also, while solving 
the coupled equations we have to express $\omega$ in the Jordan frame by using $\tilde \omega = \omega \gamma^2$ and transform all derivatives w.r.t.
$N$ to derivatives w.r.t. $\tilde N$ by using \eqref{NtotildeN}.

With this information, we solve the system of coupled equations numerically to find  the dimensionless scalar field $\varphi$ and the Hubble parameter $H$, as functions of the number of e-folds  $\tilde N$ (and the temperature). 
We choose the same initial conditions for the scalar field and its derivative as in the conformal case and to obtain the initial condition for $H$, we use \eqref{Hdis2}.

\begin{figure}[h!]
\centerline{
\includegraphics[width=.65\textwidth]{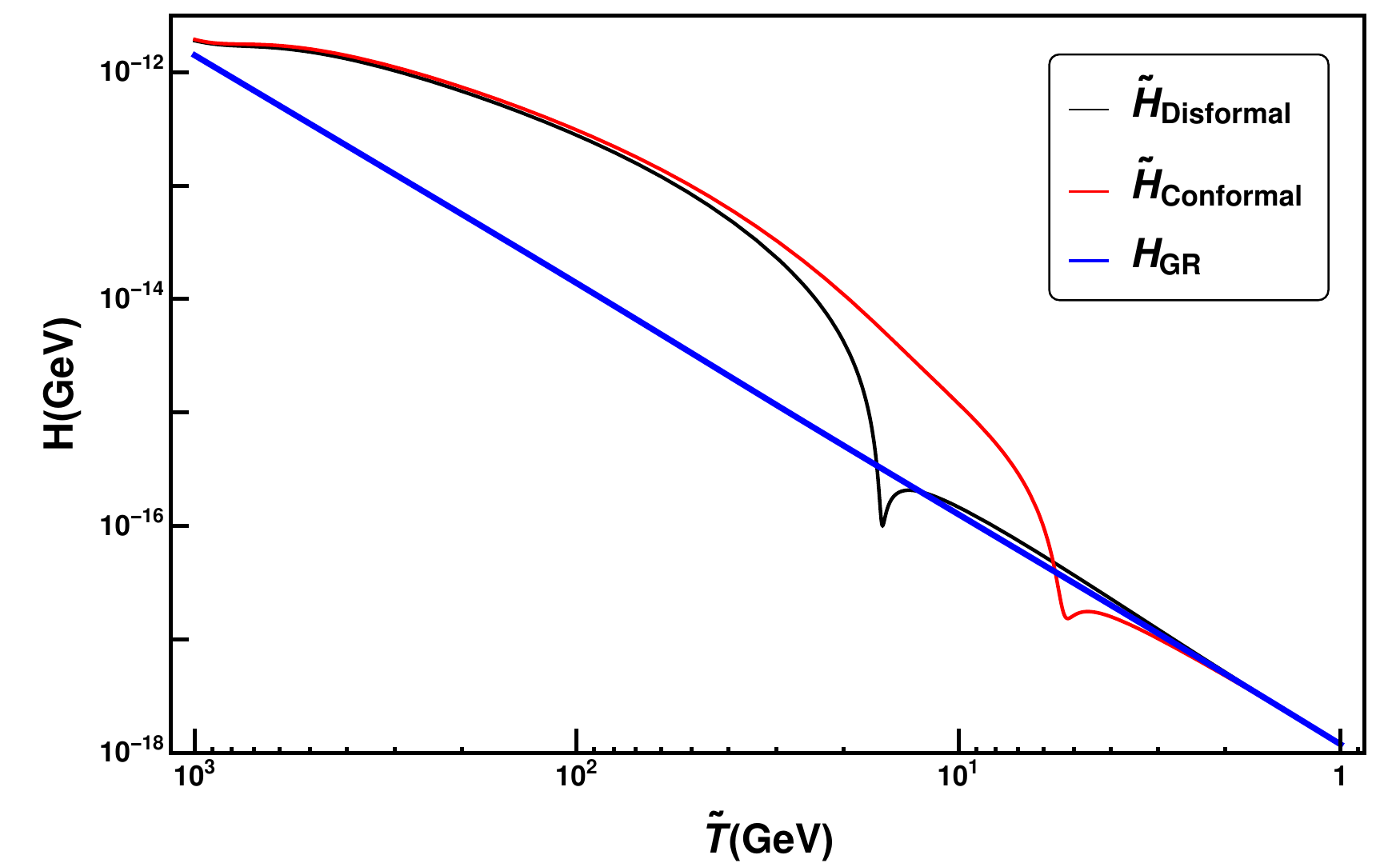}}
\caption{Comparing the modified expansion rate of the universe in the disformal and conformal scenarios for the same initial conditions as in Fig.~\ref{HubblesFig}.}
\label{Hdisformal}
\end{figure}

Once we have the solutions for $\varphi$ and $H$ as functions of temperature,  we can go back to \eqref{Htilde} and \eqref{gammaH}  to obtain the expansion rate for the disformal model. As an example, in Figure \ref{Hdisformal} we show the effects of a disformal factor given by  $D(\varphi) = D_0\,\varphi^2$ with $D_0 =-4.9\times 10^{-14}$.
In this plot, we illustrate the effect of the disformal contribution on the  expansion rate  ($\tilde H_{Disformal}$) and compare it to the modified expansion rate for the conformal case ($\tilde H_{Conformal}$) and the standard case ($H_{GR}$).
 We use the same initial conditions as in Figures \ref{FieldFig} and \ref{CFig} for the scalar field and its derivative.

 Also, it is important to mention that for the case shown the parameter constraints described in section \ref{ParCons} are satisfied. In particular we find  $\alpha_0^2 \simeq 2\times10^{-5}$, $\alpha'_0 > 0$ and $\xi\approx1.02$.

%
From our example, with $C$ and $D$ as indicated above, we can clearly see the differences from the disformally modified expansion rate $\tilde H_{Disformal}$  compared to the  conformally modified  and standard case, $H_{GR}$. The evolution of 
$\tilde H_{Disformal}$ is similar to that of  $\tilde H_{Conformal}$, having  an (two in our example) enhancement and a decrement compared to the standard expansion rate $H_{GR}$.
Moreover, the main differences with respect to the conformal modification  are the position of the notch and its shape. The notch is moved to higher temperatures and it becomes a little bit sharper.

These differences between the expansion rates can be understood from \eqref{phiHeq}. First, in this equation we see that the factor
$\frac{3H^2\gamma^2BD}{\kappa^2(1+\lambda)C}$ in the coefficients of $\varphi''$, $\varphi'$ and $\varphi'^2$ vanishes when $D=0$. For the disformal 
example shown in Figure \ref{Hdisformal},  this factor is a very small correction to the equation, which is reflected in the slight shape modification
of $\tilde H_{Disformal}$ compared to $\tilde H_{Conformal}$. Second, the term proportional to $\delta(\varphi)$ plays a more important role, being  responsible for the shifting of the notch. 

As was discussed previously in section \ref{Impact}, a modification to the expansion rate of the universe prior to BBN 
has tremendous consequences on the abundance, $\tilde Y$, of dark matter particles. Also, this modification implies that the thermally-averaged annihilation 
cross section, $\langle \sigma v \rangle$, for dark matter particles differs significantly from the one predicted by the standard cosmological model, which is 
approximately $3.0 \times 10^{-26}\rm{cm^3/s}$ (see Figure \ref{AnniFig}). We have seen than the enhancement of $\tilde H$ allows bigger values of 
$\langle \sigma v \rangle$ for particles with masses within certain range, and also, the lower value of $\tilde H$ implies smaller $\langle \sigma v \rangle$  for 
particles with masses within a small interval. Thus, the location and shape of the notch determines for which masses the annihilation cross section is smaller.
In the disformal scenario, the notch has been moved to higher temperatures, which allows particles with higher masses to have smaller and larger annihilation cross sections for the observed DM content.
We leave the detailed study of the modification of $\tilde H$ for more general conformal and disformal factors for a separate forthcoming publication.

\section{Conclusions}\label{Sec:3}

Scalar-tensor theories of gravity are a useful method to explore departures of the expansion rate of the universe from the standard cosmological model in the early universe. The expansion rate of the universe had a strong influence on the evolution of the dark matter abundance during the early stages
of the universe's evolution, specially prior to BBN. Modifications to the expansion rate during that time would be reflected in  the calculation of the dark matter relic abundance and so can be used as  a probe to the predictions of scalar-tensor theories. In this paper, we explored the role played by the scalar field in the
modification of the expansion rate of the universe on scalar-tensor theories coupled both conformally and disformally to matter.

For the conformal case, we explored a conformal factor of the type $C=(1+b\,e^{-\beta\,\varphi})^2$. 
We made no approximations  and solved numerically  the master equation for the scalar \eqref{masterConf2} for a suitable range of  initial conditions. Using this result we then computed the expansion rate modification $\tilde H$ under the presence of the scalar field during the radiation dominated era prior to BBN. When comparing the expansion rate, $\tilde H$, to the standard expansion rate, $H_{GR}$, we found that the speed-up factor, $\xi=\frac{\tilde H}{H_{GR}}$, increases up to 200 and then become of order 1 prior to BBN (see Fig.~\ref{HubblesFig}). Previously, in reference~\cite{MW}, it was shown that there is  no change in the  expansion rate arising from the modification due to $C$ compared to the standard cosmology, after satisfying all the constraints based on the boundary conditions chosen in~\cite{Catena,Catena2}.

This enhancement on the expansion rate has important consequences on the evolution of the abundance of dark matter particles. So, we also investigated the effect on the abundance of dark matter particles. We observed that for dark matter particles of large mass ($m\sim 10^3$GeV in our example, compared to the $m\sim 50$GeV reported in \cite{Catena} for which we found no re-annihilation), the particles  undergo a second 
annihilation process and then freeze-out once and for all in (see Fig.~\ref{Y1000Fig}). Moreover, we found that for large masses the annihilation cross-section has to be up to four times larger than that of  standard cosmology models in order to satisfy the dark matter content of the universe of 27 \%. On the other hand,
for small masses this re-annihilation process is not present, but we found that for masses around 130 GeV, the annihilation cross-section can be smaller than the annihilation cross-section for the standard cosmological model (see Fig.~\ref{AnniFig}).

We also started to investigate the effects on the early evolution of the expansion rate of a disformal factor in the metric \eqref{gtilde}. 
We noticed that in order to have a  consistent solution, i.e. a real positive $\tilde H$, the conformal and disformal factors  need to 
satisfy a very specific relation, \eqref{ConditionD}.  We studied the effect of a disformal factor by turning on a small disformal contribution to the conformal case, given by $D(\varphi)=-4.9\times 10^{-14}\varphi^2$. To find the modified expansion rate $\tilde H$  during the radiation dominated era prior to BBN, we solved numerically the disformal system  of coupled equations  \eqref{phiHeq} and \eqref{Hprime}. 

We found that when the disformal function is turned on, $\tilde H$  has a very similar profile as for pure conformal case with an enhancement and a notch compare to the standard 
expansion rate. However, the position of the notch changes and there may be a second enhancement of $\tilde H$ compared to $H_{GR}$, depending on the initial conditions (see also  Appendix \ref{NumImp}). The disformal factor, is moving the notch to higher temperatures which means that we can have larger and smaller annihilation cross-section for any mass of the DM candidate for the observed DM content.

We have shown the analysis of the disformal case in a concrete  example. We plan to study the disformal effects in more detail in order to asses their  general features in a forthcoming publication.  Moreover, as we mentioned at the beginning of the paper, we are aiming at models that can be embedded in a more fundamental theory of gravity such as string theory. We plan to analyse these cases in a future publication.

\acknowledgments 
We would like to thank, David Cerde\~no and Nicolao Fornengo for discussions and Gianmassimo Tasinato for discussions and comments on the manuscript.  BD and EJ acknowledge support from DOE Grant DE-FG02-13ER42020.


\begin{appendix}

\section{Numerical Implementation}\label{NumImp}

In this section we describe in detail our numerical method to study the master equation and expansion rate evolution and present additional examples for different initial conditions. 

As we explained in section \ref{ERM} the result shown in Figure \ref{plotomegaradiation} for $\tilde\omega(\tilde T)$ was obtained using the so-called ``kick'' function, defined in \eqref{kick}.
After using $\tilde \rho \simeq \pi^2 g_{eff}(\tilde T) \tilde T^4/30$, \eqref{erhoA} and \eqref{epA} $\Sigma$ becomes
\be\label{kick1}
\Sigma(\tilde T)  = \sum_{A} \frac{15}{\pi^4} \frac{g_A}{g_{eff}(\tilde T)} \,y_A^2 \int_{y_A}^{\infty} dx \frac{\sqrt{x^2-y^2_A}}{e^x\pm 1}\,,
\ee 
where $g_{eff}(\tilde T)$ is the total number of relativistic degrees of freedom\footnote{To calculate $g_{eff}$ we follow the numerical procedure described in Appendix A of \cite{Erickcek:2013dea}.}, $g_A$ is the number of internal degrees of freedom of particles of type $A$, $y_A=m_A/\tilde T$ and the plus (minus) sign in the integral is for fermions (bosons).
The particle spectrum used to compute $\Sigma$ is shown in Table \ref{spectrum} and in Figure \label{sigmafig} we plot the resulting kick function during the radiation dominated era. 
Then, we compute the equation of state from  \eqref{eqomega}   as  $\tilde\omega=(1-\Sigma(\tilde T))/3$, which is shown in Figure \ref{plotomegaradiation}.

\begin{table}
\begin{center}
 \begin{tabular}{|lll|}
 \hline
  Particle &Mass (GeV)&$g_A$\\ \hline
    $t$ &173.2 &12\\ 
    $H$ &125.1 &1\\
    $Z$ &91.19 &3\\
    $W^\pm$ &80.39 &6\\
    $b$ &4.18 &12\\
    $\tau$ &1.78 &4\\
    $c$ &1.27 &12\\
    $\pi^0$ &0.140 &12\\
    $\pi^\pm$ &0.135 &2\\
    $\mu$ &0.106 &4\\
    $e$ &0.000511 &4\\ \hline
 \end{tabular}
\end{center}
\caption{Spectrum of particles used to calculate the kick function \eqref{kick1}. For each particle, we show its mass and number of internal degrees of freedom, $g_A$.}
\label{spectrum}
\end{table}

\begin{figure}[h!]
\centerline{
\includegraphics[width=.8\textwidth]{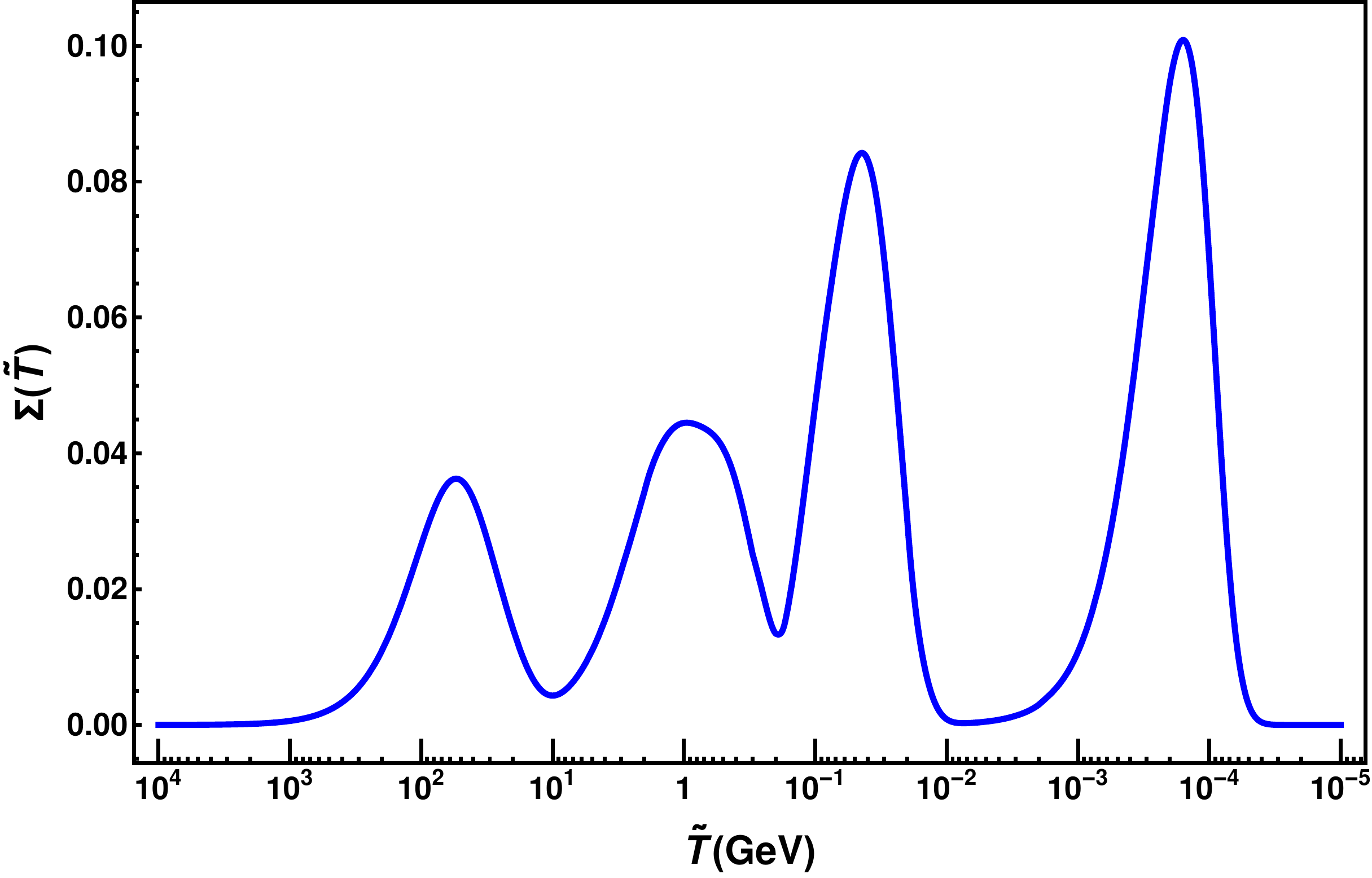}}
\caption{Thermal evolution of the kick function during the radiation dominated era. Outside the interval of temperatures shown, $\Sigma$ vanishes.}
\label{sigmafig}
\end{figure}

\subsection{Conformal case solutions}

\begin{figure}[h!]
\centerline{
\includegraphics[width=.85\textwidth]{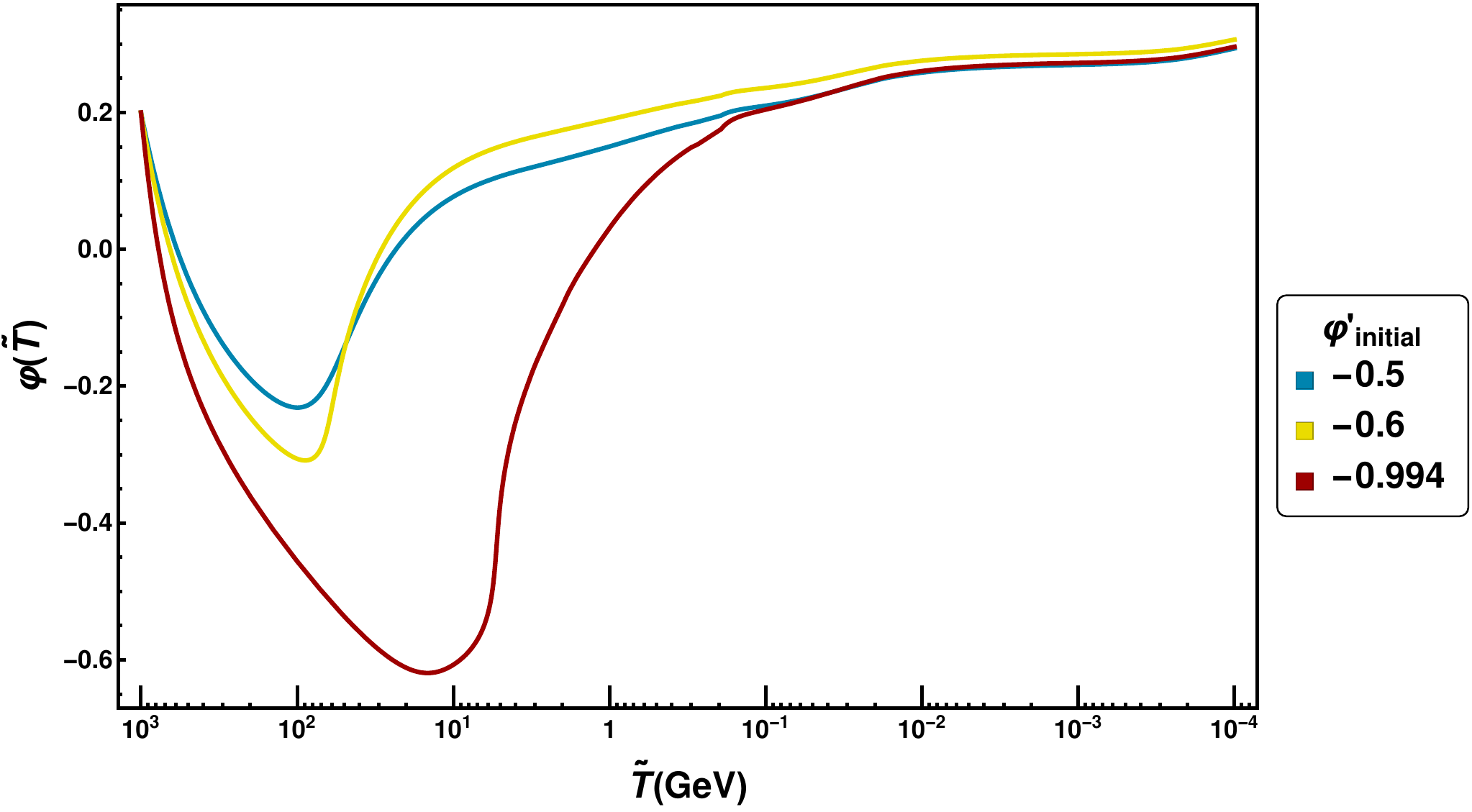}}
\caption{Scalar field as function of temperature in the pure conformal scenario for various initial conditions. We use $\varphi_{i}= 0.2$ and every curve shown 
corresponds to a different value of $\varphi'_i$.}
\label{plotsphiBC}
\end{figure}

\begin{figure}[h!]
\centerline{
\includegraphics[width=.85\textwidth]{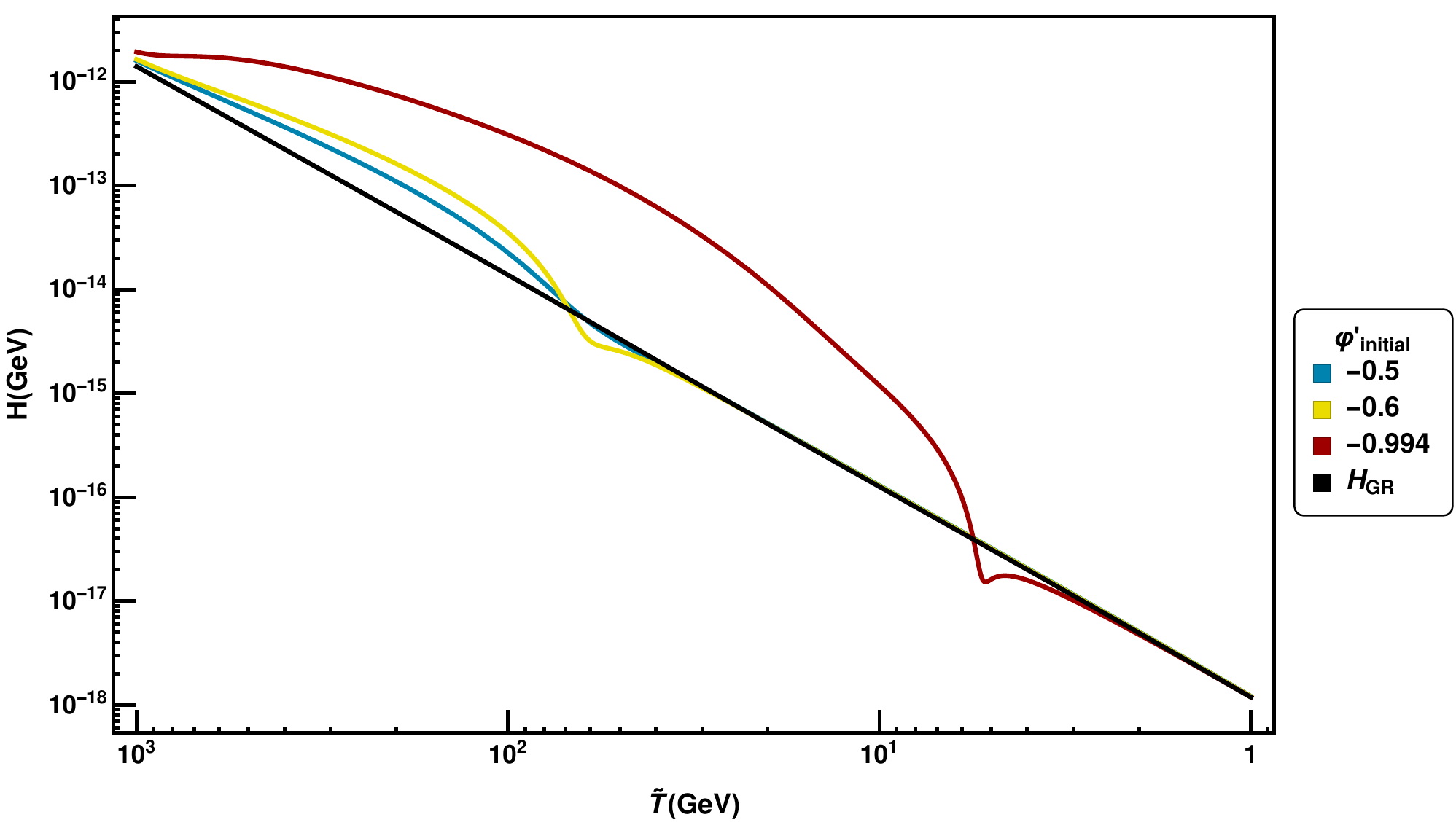}}
\caption{Modified expansion rate as function of temperature in the conformal scenario for different boundary conditions. We use $\varphi_{i}= 0.2$ and every curve shown 
corresponds to a different value of $\varphi'_i$.}
\label{plotHsconformal}
\end{figure}


Using the  numerical solution for $\tilde \omega$, we plugged it into the master equation, and solve it numerically for the interesting initial conditions as described in section \ref{ERM} for the conformal case. To look for suitable solutions, we fixed the initial value of the velocity and then look for an initial value of the scalar such that $(1+\alpha(\varphi)\varphi')$ stayed positive, thus giving a positive modified expansion rate. As we explained in that section, we were aiming at solutions where the scalar field passed from positive to negative to positive values again.
In Figure \ref{plotsphiBC}, we show  the thermal evolution of the scalar
field in the pure conformal scenario for different initial velocities $\varphi'$.  We let  the initial value of $\varphi$, fixed at $\varphi_i=0.2$ and solve  the master equation \eqref{masterConf2} for the different initial velocities and  initial temperature to be 1000 GeV. This plot shows the behaviour of the scalar field as  described in section \ref{ERM}.
In Figure  \ref{plotHsconformal}, we show the resulting  modified expansion rates for the different initial  conditions considered in Fig.~\ref{plotsphiBC}. 

\subsection{Disformal case solutions}

\begin{figure}[h!]
\centerline{
\includegraphics[width=.85\textwidth]{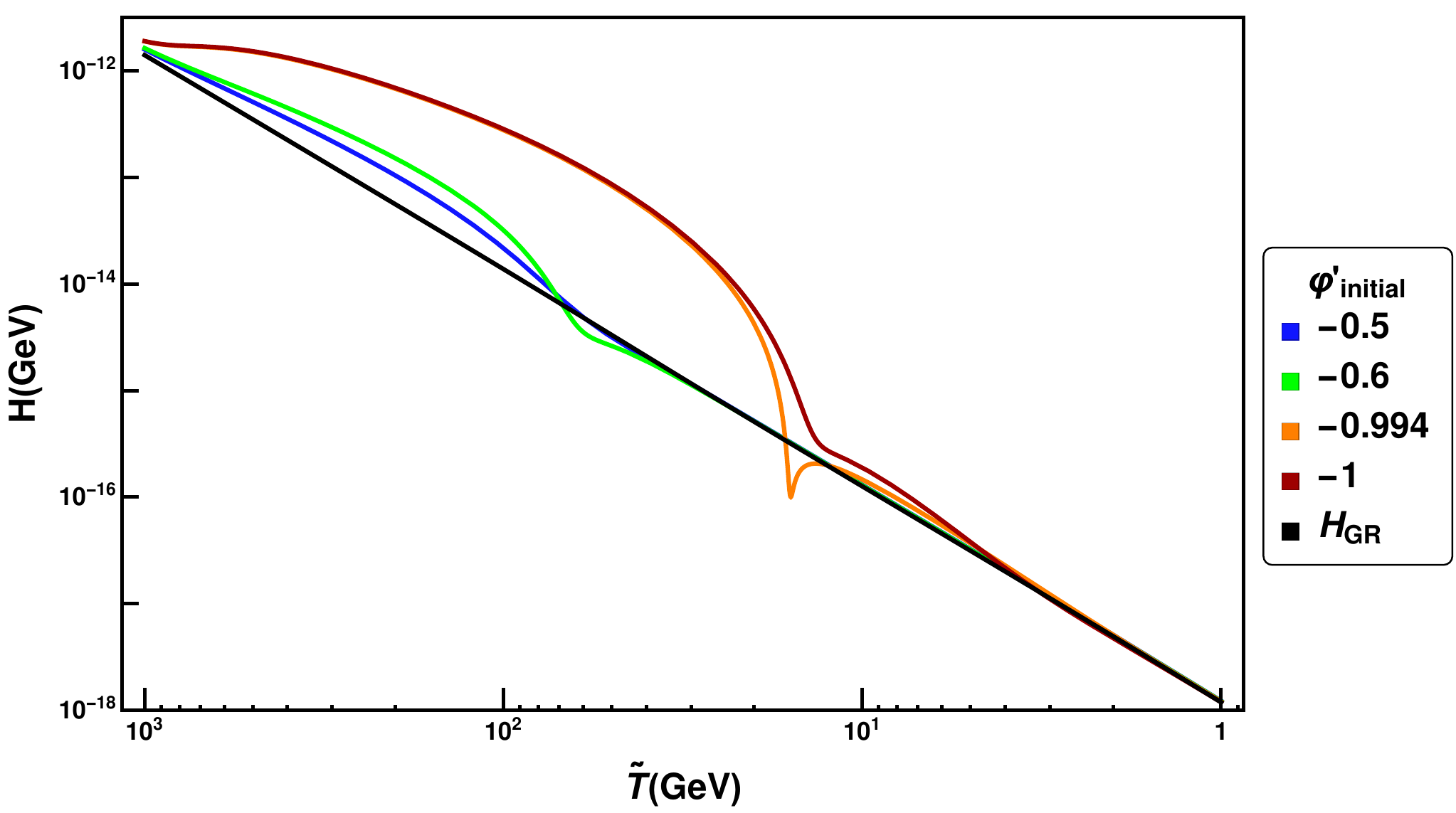}}
\caption{Modified expansion rate as function of temperature in the disformal scenario for various initial conditions.
We use $\varphi_{i}= 0.2$ and every curve shown corresponds to a different value of $\varphi'_i$}
\label{plotHsdisformal}
\end{figure}

As discussed in the main text,  when both conformal and  disformal functions are non-zero, we solve the system of coupled equations \eqref{phiHeq} and \eqref{Hprime} numerically with Mathematica. We explored different boundary conditions for the scalar field and its first derivative. With the solutions of these equations we used \eqref{Htilde} to find the modified expansion rate. In Figure \ref{plotHsdisformal} we show the modified expansion rates for a disformal factor given by $D=D_0\varphi^2$ with $D_0=-4.9\times 10^{-14}$ and the conformal factor being the same as before, that is $C= (1+b e^{-\beta\varphi})^2$ with $b=0.1$, $\beta = 8$. 
We have added one additional initial condition with respect to the conformal case, $\varphi'_{initial}=-1.$ It is interesting to notice that for this initial condition, the pure conformal case does not give a  solution satisfying the necessary constraints  explained in the main text. In this sense, we see that the disformal contribution is important in order to find solutions otherwise excluded.

\section{General Disformal Set-Up}\label{App1}

The general scalar-tensor  action coupled to matter, which can include a realisation in string theory compactifications is given by: 
\bea\label{S1A}
S=S_{EH} + S_\phi + S_{m}\,,
\eea
where: 
\bea
&& S_{EH}=\frac{1}{2\kappa^2} \!\!\int{\!d^4x\sqrt{-g}\,R}, \\
&&  S_\phi = - \!\!\int{\!d^4x\sqrt{- g} \left[\frac{b}{2} (\partial\phi)^2+  M^4 \c1 ^2(\phi)\sqrt{1+\frac{\d1(\phi)}{\c1(\phi)} (\partial\phi)^2}+V(\phi)\right]}  \,,\\
&& S_{m} = - \!\!\int{\!d^4x\sqrt{-\tilde g} \,{\cal L}_{DM}(\tilde g_{\mu\nu}) } \,,
\eea
and the disformally coupled metric is given by
\be\label{gammaA}
\tilde g_{\mu\nu} = C_2(\phi) g_{\mu\nu} + D_2(\phi) \partial_\mu \phi \partial_\nu \phi\,.
\ee
 $b$ is a constant equal to $1$ or $0$, depending on the model one wants to consider; $C_i(\phi), D_i(\phi)$ are functions of $\phi$, which can be identified as  conformal and disformal couplings of the scalar to the  metric, respectively. 
  Finally,  we have introduced the mass scale $M$  to keep units right (remember that the conformal coupling is dimensionless, whereas the disformal  has units of $Mass^{-4}$.)

The connection of the general action \eqref{S1A} to the different models in the literature can be obtained as follows:  the case $C_1 = D_2$, $D_1=D_2$, $b=0$ arises when considering a D-brane moving along an extra dimension. This case was  studied in \cite{KWZ} as a model of a coupled dark matter dark energy sector scenario, where scaling solutions arise naturally.  
Note that in this case, the kinetic term for the scalar field, identified with dark energy for example, is automatically  non-canonical and  dictated by the DBI action (see \cite{KWZ}). 
On the other hand, phenomenological models considering a disformal coupling between matter and a scalar field, usually consider a canonical kinetic term, and therefore, in that case, $C_1=D_1=0$ and $b=1$ ($C_1$ can be taken to be non-zero and will be part of the scalar potential). 
Furthermore, the widely studied case of a conformal coupling is obtained for $b=1, C_1=D_1=D_2=0$ or, as in the case of a D-brane for example\footnote{For the system corresponding to a D-brane moving in a typically warped compactification in string theory, the functions $C(\phi)$ and $D(\phi)$ are identified with powers of the so-called warp factor, usually denoted as $h(\phi)$. In this approach,  the longitudinal and transverse fluctuations of the D-brane are identified with the dark matter and dark energy fluids respectively \cite{KWZ}.}, simply considering small velocities with   $b=0, C_1=C_2$ and $D_1=D_2$, and normalising canonically the scalar field (see Appendix \ref{App2}). 

Finally, let us clarify further our nomenclature on frames.  The action in \eqref{S1A} is written in the Einstein frame, which  in string theory, is usually related to the frame in which the dilaton and the graviton degrees of freedom are decoupled. From this point of view, the dilaton field as well as all other moduli fields not relevant for the cosmological discussion are considered as stabilised, massive, and are therefore decoupled from the low energy effective theory. 
In the literature of  scalar-tensor theories however (including conformal and disformal couplings), the Einstein and Jordan frames are identified with respect to the (usually single) scalar field to which gravity is coupled. In this paper, we follow this and call  ``Jordan" or ``disformal frame"  the frame in which dark matter is coupled only to the metric $\tilde g_{\mu\nu}$, rather than to the metric $g_{\mu\nu} $ and a scalar field $\phi$.


The equations of motion obtained from  (\ref{S1A}) are \eqref{EM}: 
\be
R_{\mu\nu} -\frac{1}{2}g_{\mu\nu} R = \kappa^2\left(T^\phi_{\mu\nu} + T^{DM}_{\mu\nu}\right)\,,
\ee
where in the  frame  relative to $g_{\mu\nu}$ the energy momentum tensors are defined in \eqref{EM1} and \eqref{EM2}. 
 The energy-momentum  tensor for the scalar field in the general case  is modified from \eqref{EMphi} to:
\be\label{EMphiA}
T_{\mu\nu}^{\phi} = - g_{\mu\nu} \left[M^4C_1^2 \gamma_1^{-1} + \frac{b}{2} (\partial\phi)^2+ V \right] 
+ \left(M^4 C_1D_1 \,\gamma_1+ b\right) \partial_\mu\phi \, \partial_\nu \phi
\ee
where now the energy density and pressure are given by:
\bea\label{rhoPA}
\rho_\phi =- \frac{b}{2}(\partial\phi)^2 + M^4C_1^2 \gamma_1 + V  \,, \qquad P_\phi = - \frac{b}{2}(\partial\phi)^2 - M^4C_1^2\gamma_1^{-1} - V  \,,
\eea
 and the  ``Lorentz factor"  $\gamma_1$ introduced above is defined by
\be\label{LorentzA}
\gamma_1 \equiv \left(1+ \frac{D_1}{C_1}\, (\partial\phi)^2\right)^{-1/2}\,.
\ee
We can rewrite \eqref{rhoPA} in a more succinct way, by defining ${\mathcal V} \equiv V + C_1^2 M^4$ 
\be\label{rhoPA2}
\rho_\phi =- \left [\frac{b}{2}+  \frac{M^4C_1 D_1\gamma_1}{\gamma+1}\right] (\partial\phi)^2 + {\mathcal V}  \,, \qquad 
P_\phi = - \left [\frac{b}{2}+  \frac{M^4C_1 D_1\gamma_1^{-1}}{\gamma+1}\right] (\partial\phi)^2 - {\mathcal V}  \,.
\ee

The equation of motion for the scalar field becomes (compare with \eqref{Eqphi})
\bea\label{EqphiA}
&&\hskip-1cm - \nabla_\mu\!\left[(M^4 D_1C_1\gamma_1\, +b)\,\partial^\mu \phi\right ] \!+ \!\frac{\gamma_1^{-1}M^4 C_1^2}{2} \!\left[\!\frac{D_1'}{D_1} +3\frac{C_1'}{C_1} \right] \!+ \!\frac{\gamma_1 \,M^4 C_1^2}{2} \!\left[\!\frac{C_1'}{C_1} -\frac{D_1'}{D_1} \right] \!+\! V' 
\nonumber \\&& \hskip3cm 
- \frac{T^{\mu\nu}}{2}\!\left[\frac{C_2'}{C_2} g_{\mu\nu}  +\frac{D_2'}{C_2}\partial_\mu\phi\partial_\nu\phi\right]
+\nabla_\mu \left[\frac{D_2}{C_2}T^{\mu\nu} \partial_\nu\phi \right] =0\,. \nonumber \\
\eea

Finally, the energy-momentum conservation equation gives rise to \eqref{conserva}, where $Q$ now is given in terms of $C_2, D_2$:
\be
Q\equiv \nabla_\mu \left[\frac{D_2}{C_2} \,T^{\mu\lambda} \,\partial_\lambda \phi\right] - \frac{T^{\mu\nu} }{2} \left[\frac{C_2'}{C_2} g_{\mu\nu} +
\frac{D_2'}{C_2} \,\partial_\mu\phi \,\partial_\nu\phi\right]\,.
\ee


\subsection{General cosmological equations}

The equations of motion for the general system in an FRW background  become:
\bea
&& H^2 =\frac{\kappa^2}{3} \left[\rho_\phi +\rho\right]\,, \label{friedmann1A}\\
&& \dot H + H^2 = -\frac{\kappa^2}{6}\left[ \rho_\phi+ 3P_\phi +\rho +3 P \right]\,,\label{friedmann2A}\\
&& \ddot \phi \left[1+ \frac{b}{M^4C_1D_1\gamma_1^3}\right]+3H\dot\phi \,\gamma_1^{-2}\left[\frac{b}{M^4C_1D_1\gamma_1}+ 1\right]  \nonumber \\
&&\hskip0.5cm + \frac{C_1}{2D_1}\left(\gamma_1^{-2}\left[\frac{5 C_1'}{C_1}  - \frac{D_1'}{D_1}\right] 
 + \frac{D_1'}{D_1}- \frac{C_1'}{C_1} -4\gamma^{-3}_1 \frac{C'}{C}  \right)
 + \frac{1}{M^4C_1D_1\gamma_1^{3}}\, ({\mathcal V}'+Q_0) =0 \label{kgA} \,,  \nonumber \\
\eea
where, $H= \frac{\dot a}{a}$, dots are derivatives with respect to $t$, $'=d/d\phi$ and  $$\gamma_1= (1-D_1 \,\dot\phi^2/C_1)^{-1/2}.$$
We also have the continuity equations for the scalar field and matter given by
\bea 
&&\dot\rho_\phi + 3H(\rho_{\phi}+P_{\phi}) = -Q_0\dot\phi\,, \label{contA}\\
&&\dot\rho + 3H(\rho+P) = Q_0\,\dot\phi\,.\label{cont1A}
\eea
where $Q_0$ is given by
\bea
Q_0 = \rho \left[ \frac{D_2}{C_2} \,\ddot \phi + \frac{D_2}{C_2} \,\dot \phi \left(\!3H + \frac{\dot \rho}{\rho} \right) \!+ \!\left(\!\frac{D_2'}{2C_2}-\frac{D_2}{C_2}\frac{C_2'}{C_2}\!\right) \dot\phi^2 +\frac{C_2'}{2\,C_2} (1-3\,\omega)
\right]. \nonumber \\
\eea
Using \eqref{cont1A} we can rewrite this in a more compact  and useful form as 
\be\label{Q0A}
Q_0 = \rho\left( \frac{\dot \gamma_2}{\dot \phi\, \gamma_2} + \frac{C_2'}{2C_2}  (1-3\,\omega \,\gamma_2^2) -3H\omega\,\frac{(\gamma_2 -1) }{\dot \phi}\right) \,,
\ee
where $$\gamma_2= (1-D_2 \,\dot\phi^2/C_2)^{-1/2}.$$

Plugging this into  the  
(non-)conservation equation for dark matter \eqref{cont1A},  gives:
\be\label{conservaDMA}
\dot \rho + 3H (\rho + P\,\gamma_2^{2}) = \rho \left[\frac{\dot \gamma_2}{\gamma_2} + \frac{C_2' }{2C_2} \,\dot\phi\, (1-3\,\omega \gamma_2^2)\right]\,.
\ee

The energy densities and pressures in the Einstein and Jordan frames are now related similarly to \eqref{eqrhos}, replacing $\gamma \to \gamma_2$:
\be\label{eqrhosA}
\tilde \rho = C_2^{-2} \gamma_2^{-1} \rho \,, \qquad \tilde P = C_2^{-2}\gamma_2\, P,
\ee
and therefore the equation of states in both frames are related by $\tilde \omega = \omega \gamma_2^2$.
Similarly  the physical proper time and the scale factors in the two frames are related via $\gamma_2$:
\be\label{tildeaA}
\tilde a = C_2^{1/2} a   \,, \qquad \quad d\tilde \tau = C_2^{1/2} \gamma_2^{-1} d\tau \,.
\ee
Defining  the disformal frame Hubble parameter $\tilde H \equiv \frac{d \ln{\tilde a}}{d\tilde \tau}$, gives:
\be\label{tildeHA}
\tilde H = \frac{\gamma_2}{C_2^{1/2}}\left[ H + \frac{C_2'}{2C_2}\dot \phi \right]\,.
\ee
To  solve the equations of motion one now can proceed as in section \ref{ME} to write the equations in terms of derivatives w.r.t.~the number of e-folds $N$ and consider different cases by choosing appropriately the parameters $b,C_i,D_i$. 
We leave the analysis of these for a future publication.

\section{The conformal case in D-brane scenarios}\label{App2}

In this section we show how to recover the pure conformal case from the D-brane picture, that is, $b=0$, $C_1=C_2$, $D_1=D_2$. 
We start by  expanding the square root in the scalar part of the action \eqref{S1A}. Doing this we get
\bea\label{Sphiconf}
S_\phi& =& -\int{d^4x  \sqrt{-g} \left[  M^4C_1^2 \left( 1+ \frac{D_1}{2C_1} (\partial \phi)^2   + \dots \right) + V(\phi)\right]}\nonumber \\
&=& -\int{d^4x  \sqrt{-g} \left[  \frac{M^4C_1 D_1}{2} (\partial \phi)^2   + M^4C_1^2(\phi)+ V(\phi) + \dots \right]}\nonumber \\ 
&=& -\int{d^4x  \sqrt{-g} \left[  \frac{M^4C_1 D_1}{2} (\partial \phi)^2   + {\mathcal V}(\phi) + \dots \right]} \,, 
\eea
On the other hand, the matter Lagrangean  takes the form
\bea
S_{DM} &=& -\int{d^4x  \sqrt{-\tilde g} \,{\cal L}_{DM}(\tilde g_{\mu\nu})}\nonumber \\
 &=& -\int{d^4x  \sqrt{-g} \, C_1^2(\phi) \left(1+ \frac{D_1}{2C_1} (\partial \phi)^2   + \dots  \right){\cal L}_{DM}( \tilde g_{\mu\nu})} \nonumber \\  &=&  -\int{d^4x  \sqrt{- g} \, C_1^2 (\phi) {\cal L}_{DM}( \tilde g_{\mu\nu}) + \dots } =-\int{d^4x  \sqrt{- \tilde g} \, {\cal L}_{DM}( \tilde g_{\mu\nu}) + \dots } \nonumber \\ 
 \eea
where now $\tilde g_{\mu\nu} = C_1(\phi) g_{\mu\nu}$ (and we have used that $\det \tilde g_{\mu\nu} = C_1^4 (1+ D_1/C_1 (\partial\phi)^2)$). 

Finally, to compare the D-brane case with the pure conformal case, we need to canonically normalise $\phi$. Calling $\varphi$ the canonically normalised field, this is obtained from $\phi$ as
\be
\varphi = \int M^2\sqrt{D_1C_1} \, d\phi \,.
\ee

It is clear that when $D_1=1/(M^4 C_1)$, $\varphi =\phi$ and therefore the action for the scalar field \eqref{Sphiconf} is already in the required form. 
We can now take the limit $\gamma \to 1$ into the  equations of motion \eqref{friedmann1A}-\eqref{kgA} and make  the identification $D_1=1/(M^4C_1)$ to recover the conformal case equations of motion. Note that in this limit $Q_0 \to \rho C_1'/2C_1$, and  is independent of $D_1$ (see \eqref{Q0A} with $C_1=C_2, D_1=D_2$).

\end{appendix}

\bibliographystyle{utphys}
\bibliography{refs}

\end{document}